\documentclass[twocolumn,conference,letterpaper]{IEEEtran}
\usepackage[left=0.75in,right=0.75in,top=1in,bottom=0.75in]{geometry}

\usepackage{cite}
\usepackage{amssymb,amsmath,latexsym,amsfonts,amsthm,mathtools}

\DeclareMathOperator*{\argmin}{arg\,min}

\usepackage{accents}
\newcommand\munderbar[1]{%
	\underaccent{\bar}{#1}}

\usepackage{algorithmic}
\usepackage{graphicx}
\usepackage{float,subcaption}
\usepackage{textcomp}
\usepackage{xcolor}
\definecolor{darkpastelpurple}{rgb}{0.59, 0.44, 0.84}
\usepackage{hyperref}
\hypersetup{colorlinks, breaklinks, citecolor=blue, linkcolor=blue, urlcolor=blue}
\usepackage[nameinlink]{cleveref}
\Crefformat{figure}{#2Fig.~#1#3}
\Crefmultiformat{figure}{Figs.~#2#1#3}{ and~#2#1#3}{, #2#1#3}{ and~#2#1#3}

\usepackage{tikz}
\usetikzlibrary{automata, shapes, arrows, calc, arrows.meta, fit, positioning}
\usepackage{pgfplots}

\theoremstyle{plain}
\newtheorem{theorem}{Theorem}
\newtheorem{lemma}{Lemma}

\newtheorem*{problem*}{Problem}
\theoremstyle{remark}

\newtheorem{assumption}{Assumption}

\newtheorem{definition}{Definition}
\theoremstyle{definition}

\newcommand{\sign}{\mathrm{sign}}
\usepackage{mathrsfs}
\newcommand{\ith}{$i$\textsuperscript{th} }
\newcommand{\M}{_{\mathrm{M}_i}}
\newcommand{\los}{_{\mathrm{L}_i}}

\usepackage{newtxmath}
\usepackage{booktabs}

\IEEEoverridecommandlockouts

\begin{document}
	\title{Cooperative Target Capture in 3D Engagements over Switched Dynamic Graphs}

\author{Abhinav Sinha,  and Shashi Ranjan Kumar

	\thanks{A. Sinha is with the Guidance, Autonomy, Learning, and Control for Intelligent Systems (GALACxIS) Lab, Department of Aerospace Engineering and Engineering Mechanics, University of Cincinnati, OH 45221, USA. (e-mail: abhinav.sinha@uc.edu).\newline S. R. Kumar is with the Intelligent Systems \& Control (ISaC) Lab, Department of Aerospace Engineering, Indian Institute of Technology Bombay, Mumbai 400076, India. (e-mail: srk@aero.iitb.ac.in).}
    }

	\maketitle
    \thispagestyle{empty}
	
	\begin{abstract}
		This paper presents a leaderless cooperative guidance strategy for simultaneous time-constrained interception of a stationary target when the interceptors exchange information over switched dynamic graphs. We specifically focus on scenarios when the interceptors lack radial acceleration capabilities, relying solely on their lateral acceleration components. This consideration aligns with their inherent kinematic turn constraints. The proposed strategy explicitly addresses the complexities of coupled 3D engagements, thereby mitigating performance degradation that typically arises when the pitch and yaw channels are decoupled into two separate, mutually orthogonal planar engagements. Moreover, our formulation incorporates modeling uncertainties associated with the time-to-go estimation into the derivation of cooperative guidance commands to ensure robustness against inaccuracies in dynamic engagement scenarios. To optimize control efficiency, we analytically derive the lateral acceleration components in the orthogonal pitch and yaw channels by solving an instantaneous optimization problem, subject to an affine constraint. We show that the proposed cooperative guidance commands guarantee consensus in time-to-go values within a predefined time, which can be prescribed as a design parameter, regardless of the interceptors' initial configurations. We provide simulations to attest to the efficacy of the proposed method.
	\end{abstract}
	
	
	\section{Introduction}\label{sec:introduction}
    C{\scshape ooperative} time-constrained guidance or cooperative salvo is crucial in both defense and civilian applications to enhance precision, efficiency, and safety in autonomous vehicles' operations. In defense, a cooperative salvo can serve as a counter/counter-countermeasure \cite{10945687,9266104,9409660,9274339,Sinha2022jint,9000526} by overwhelming the adversary's defenses via simultaneous or sequential arrival at the target. On the other hand, non-defense applications of salvo guidance \cite{10271366,4476150,9800919} span space exploration, disaster response, precision agriculture, autonomous robotics, etc.

    The development of sophisticated guidance strategies for a single interceptor operating under terminal time constraints for target capture has laid the groundwork for the creation of advanced guidance strategies \cite{1597196,Sinha2022jint,10945687,9266104,SINHA2021106824,SINHA2021106776}. These approaches augment a baseline guidance command with a feedback control term to address the terminal constraints. For instance,  \cite{Jeon2010,CHEN20195692,doi:10.2514/1.G005367}, and references therein have successfully applied this concept to design guidance commands for a swarm of interceptors against a single target, thereby demonstrating the versatility of this methodology. 

    Given the inherently short duration of engagements, it is critical for interceptors to rapidly converge on key quantities of interest, particularly in scenarios requiring simultaneous target capture. This necessitates precise control over both the engagement duration and the time required for consensus among interceptors on these critical parameters. Typically, regulating an interceptor’s time-to-go, which is a metric representing its remaining engagement duration \cite{Jeon2010}, enables trajectory shaping to satisfy time-constrained interception requirements. As a result, time-to-go serves as a fundamental coordination variable \cite{doi:10.2514/1.G005367}, facilitating simultaneous target capture. 
    
    Previously, cooperative guidance strategies with asymptotic \cite{9147677,LYU2019100,11018241}, finite-time \cite{9000526,9266104,SONG2017193}, and fixed-time \cite{9540264,CHEN2021106523} protocols have been utilized for consensus in time-to-go values. In scenarios where initial engagement parameters are either unavailable or highly uncertain, predicting the consensus time prior to homing becomes challenging. In the worst case, the consensus time (in the case of asymptotic or finite-time methods) may exceed the total engagement duration, leading to degraded guidance accuracy and potential mission failure. Fixed-time consensus mitigates this issue by ensuring that consensus/convergence time is independent of initial conditions, thereby enhancing the guidance precision. However, achieving a specific consensus time often requires conservative controller gain selection, potentially leading to an overestimation. Consequently, the design parameters may need iterative re-tuning to accommodate varying consensus time requirements effectively.

    It is also important to note that the majority of existing cooperative guidance strategies, see, for example, \cite{1597196,Sinha2022jint,10945687,9266104,SINHA2021106824,SINHA2021106776} and references therein, assume a fixed interaction network among interceptors. Throughout the engagement, a given interceptor, say $I_1$, consistently exchanges time-to-go information with the same set of neighboring interceptors. In cooperative simultaneous target capture scenarios, interceptors may coordinate by exchanging estimates of their time-to-go (or expected target capture time) via a shared communication network. However, in practical settings, the ability of each interceptor to acquire and transmit the necessary engagement variables is constrained by sensing limitations and communication network dynamics. Consequently, interactions among interceptors are not static. An interceptor’s neighbors may change over the course of the engagement. For cooperative guidance laws to remain effective under such constraints, they must be designed to accommodate switching network topologies, ensuring robust coordination despite dynamic communication links. There exists a limited body of research on cooperative guidance strategies that account for interceptors' switching topologies (see, for instance, \cite{SINHA2022107686,ZHANG2020105641,Zhao2017NoDy,Zhao2019NoDy,SUN20141202,ZHAO20171570}), which address 2D engagement scenarios.  

    However, consideration of a 3D scenario is much more challenging than a 2D engagement due to the fact that 3D engagements exhibit strong cross-coupling in their yaw and pitch channels, and control allocation complexity becomes pronounced (which is otherwise absent in a 2D setting), thereby making it nontrivial to determine the optimal lateral acceleration components when the interaction within the swarm is dynamic. There have been recent interest in designing 3D cooperative guidance with inherent cross-coupling (see, for instance, \cite{10945687,doi:10.2514/1.G005367,9409660}), but the design is mostly limited to static networks. Note that in 3D engagements, the challenge of designing a robust cooperative guidance strategy is greater under continually changing interaction topologies and the increased complexity of the nature of engagement. In this paper, we aim to address this challenge, where our focus is specifically on
    \begin{itemize}
        \item designing a cooperative guidance strategy for a swarm of interceptors to capture a target in 3D when the interceptors communicate over dynamic graphs,
        \item being able to exercise control over the time needed by the interceptors to agree on a common time-to-go,
        \item ensuring robustness to modeling uncertainties in the estimation of time-to-go, and
        \item allowing efficient allocation of the net control authority (the lateral acceleration) in the pitch and the yaw channels
    \end{itemize}
    in the absence of the radial acceleration component, to guarantee a simultaneous target capture. We further analyze the dynamics of necessary engagement variables to provide further insights into the cooperative behavior. 
    
    The remainder of this paper is organized as follows. After an introduction in \Cref{sec:introduction}, necessary background and problem formulation are presented in \Cref{sec:problemformulation}. Derivations of the cooperative guidance command are presented in \Cref{sec:st} while simulations are shown in \Cref{sec:simulations}. Finally, \Cref{sec:conclusions} concludes the paper and discusses some outlook towards future investigations.
	
\section{Background and Problem Formulation}\label{sec:problemformulation}
\begin{figure}[ht!]
		\centering
		\begin{subfigure}[t]{\linewidth}
			\includegraphics[width=\linewidth]{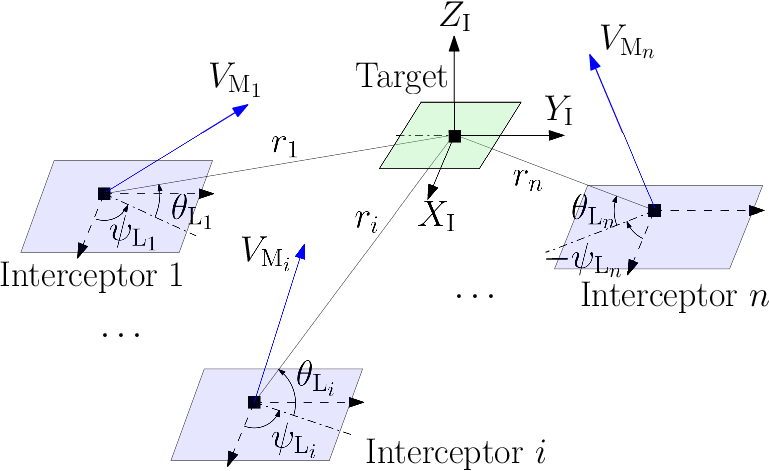}
			\caption{Multi-interceptor homing on to a single target.}
			\label{fig:eng_geo_multi}
		\end{subfigure}
		\begin{subfigure}[t]{\linewidth}
			\includegraphics[width=\linewidth]{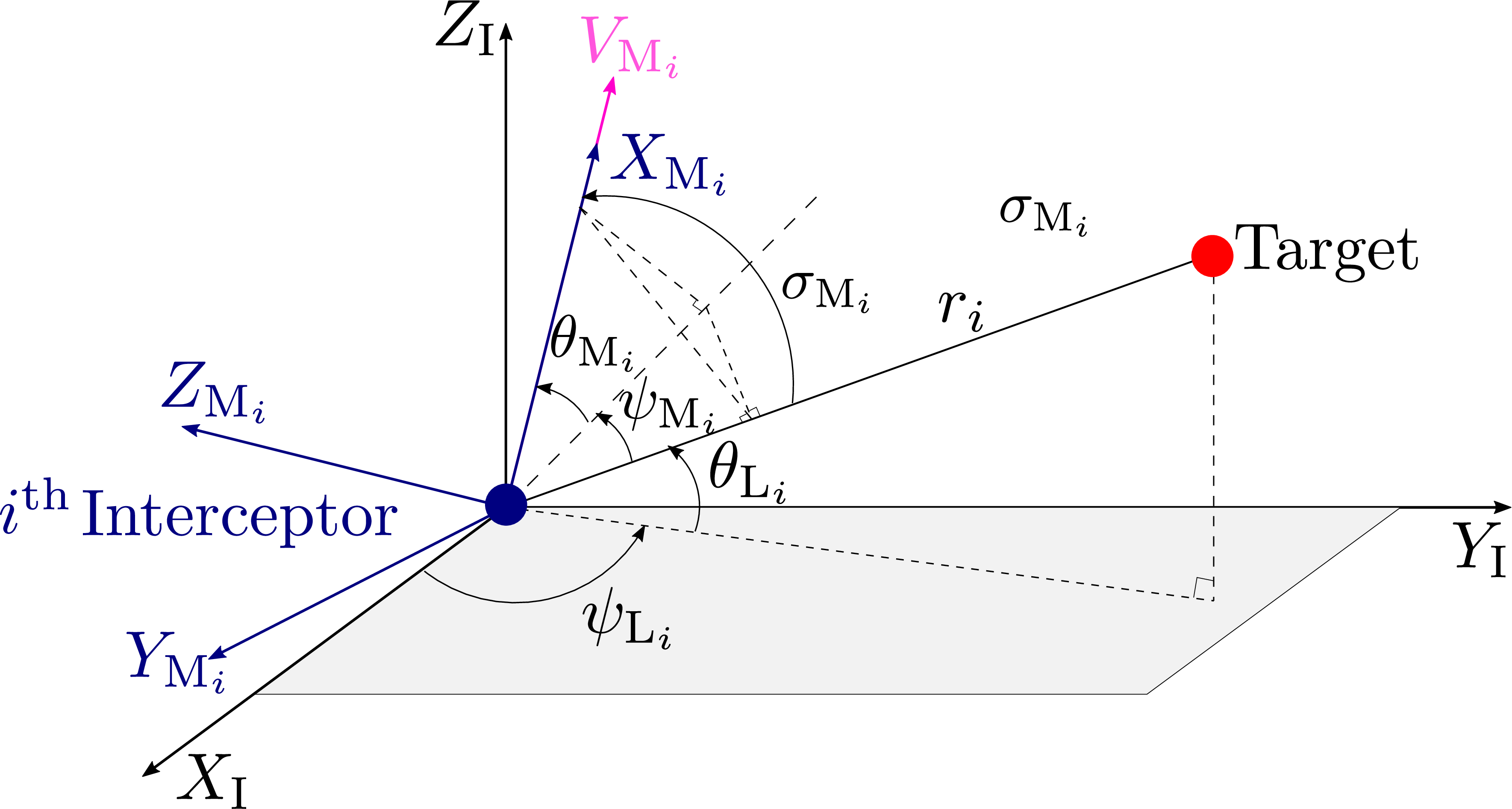}
\caption{Engagement for the $i$\textsuperscript{th} interceptor.}
\label{fig:eng_geo_3D}
		\end{subfigure}
		\caption{Cooperative target engagement in 3D.}
  \label{fig:enggeo}
	\end{figure}

Consider a three-dimensional engagement scenario involving multiple interceptors and a stationary target, as illustrated in \Cref{fig:enggeo}. In this context, each interceptor is indexed using a subscript $\mathrm{M}_i$, while the target is denoted by $\mathrm{T}$. The engagement takes place within an inertial frame of reference defined by the mutually orthogonal axes ${X_\mathrm{I},Y_\mathrm{I},Z_\mathrm{I}}$. The relative range between the $i$\textsuperscript{th} interceptor and the target is represented by $r_i$, and the line-of-sight (LOS) between them is characterized by the azimuth angle $\psi_{\mathrm{L}_i}$ and elevation angle $\theta_{\mathrm{L}_i}$, both measured with respect to the $X_\mathrm{I}$-axis.
    The LOS coordinate frame for each interceptor is defined as ${X_{\mathrm{L}_i},Y_{\mathrm{L}_i},Z_{\mathrm{L}_i}}$, where $X_{\mathrm{L}_i}$ points along the LOS vector connecting the interceptor and the target, $Y_{\mathrm{L}_i}$ is oriented $90^\circ$ counterclockwise from $X_{\mathrm{L}_i}$, and $Z_{\mathrm{L}_i}$ completes the right-handed coordinate system. In contrast, the body-fixed reference frames of the vehicles are denoted by ${X_{\mathrm{M}_i},Y_{\mathrm{M}_i},Z_{\mathrm{M}_i}}$ for the interceptors and ${X_\mathrm{T},Y_\mathrm{T},Z_\mathrm{T}}$ for the target. The velocity vector of the interceptor is aligned with its body-fixed $X$-axes and is expressed as $\mathbf{V}_{\mathrm{M}_i} = [V_{\mathrm{M}_i};0;0]^\top$. It is assumed that the target is stationary. The orientation of the $X_{\mathrm{M}_i}$ axis within the LOS frame is determined by the lead angles ${\psi_{\mathrm{M}_i}, \theta_{\mathrm{M}_i}}$.

Unlike approaches that utilize both radial and lateral acceleration components, e.g. \cite{9761742}, we assume that the vehicles are capable of maneuvering solely via lateral acceleration, without access to radial acceleration. This constraint restricts their control authority, resulting in trajectories governed exclusively by the lateral acceleration components while maintaining constant speeds throughout the engagement. This assumption is motivated by the practical challenges of implementing controlled thrust and the relatively minor variations in interceptor speed during the terminal phase of an engagement. Accordingly, the lateral acceleration components of the \ith interceptor are denoted as $a_{\mathrm{M}_i}^z$ and $a_{\mathrm{M}_i}^y$. Here, $a_{\mathrm{M}_i}^z$ acts along the $Z_{\mathrm{M}_i}$ axis and influences the interceptor’s pitch (elevation) in the LOS frame, while $a_{\mathrm{M}_i}^y$ acts along the $Y_{\mathrm{M}_i}$ axis and adjusts the azimuth, enabling coordinated simultaneous interception, which is central to the objective of this study.

Based on the aforementioned notations, we can now describe the kinematic equations governing the relative engagement between the $i$\textsuperscript{th} interceptor and the target in the spherical coordinate system, as
\begin{subequations}\label{eq:engkinematics}
		\begin{align}
			\dot {r}_i&=-V\M\cos{\theta\M}\cos{\psi\M},\label{eq:rdot}\\
			\dot{\theta}\los&=-\dfrac{V\M\sin{\theta\M}}{r_i},\label{eq:thetadot}\\
			\dot{\psi}\los&=-\dfrac{V\M\cos\theta\M\sin\psi\M}{r_i\cos\theta\los},\label{eq:psidot}\\ 
			\dot{\theta}\M&=\dfrac{a\M^z}{V\M}-\dot{\psi}\los\sin\theta\los\sin\psi\M-\dot{\theta}\los\cos\psi\M,\label{eq:thetaMdot}\\
			\dot{\psi}\M&=\dfrac{a\M^y}{V\M\cos\theta\M}+\dot{\psi}\los\tan\theta\M\cos\psi\M\sin\theta\los\nonumber\\
            &-\dot{\psi}\los\cos\theta\los-\dot{\theta}\los\tan\theta\M\sin\psi\M.\label{eq:psiMdot}
		\end{align}
	\end{subequations}
From \eqref{eq:engkinematics}, it is evident that there is a strong coupling between the two planes. To characterize this relationship, we introduce the concept of \textit{effective lead angle} for the \ith interceptor, denoted as $\sigma\M$ determined based on the azimuth and elevation angles in the LOS frame. Therefore,
\begin{align}
\cos\sigma\M =& \cos\psi\M\cos\theta\M.\label{eq:sigmaM}
\end{align}
In modern defense and autonomous systems, the ability to coordinate a swarm of interceptors for the simultaneous capture of a stationary target in 3D settings is of significant operational value. This objective becomes considerably more complex when the communication topology among interceptors is dynamic, as is often the case in real-world scenarios where communication links are affected by range limitations, environmental occlusions, or strategic disruptions. Designing cooperative guidance strategies that guarantee simultaneous interception under such dynamic conditions is both a theoretically rich and practically relevant challenge. We now present the main problem addressed in this paper.
\begin{problem*}
Design a cooperative guidance strategy to ensure that a swarm of interceptors simultaneously captures a stationary target in 3D settings when their interaction topology is dynamic. 
\end{problem*} 
The cooperative interaction among multiple interceptors can be effectively represented using graph-theoretic models. A graph is defined as an ordered pair $\mathcal{G} = (\mathcal{V}, \mathcal{E})$, where $\mathcal{V} = \left\{1, 2, \ldots, n\right\}$ is a finite set of vertices representing the $n$ interceptors, and $\mathcal{E} \subset \left\{i,j \mid i, j \in \mathcal{V}; i \neq j \right\}$ is the set of edges corresponding to the communication links between them. Unless stated otherwise, the number of edges is denoted by $E$. The neighborhood of a vertex $i$, denoted by $\mathcal{N}_i$, is defined as the set of vertices connected to $i$ by an edge, that is, $\mathcal{N}_i = \left\{j \in \mathcal{V} \mid {i,j} \in \mathcal{E}\right\}$. For an undirected graph with vertex set $\mathcal{V}$, the associated Laplacian matrix $\mathcal{L}(\mathcal{G})$ has entries $l_{ij}$ defined as  $l_{ij} = -1$ if ${i,j} \in \mathcal{E}$, $l_{ii} = |\mathcal{N}_i|$ (i.e., the degree of node $i$), and $l_{ij} = 0$ otherwise. The Laplacian matrix is positive semi-definite, and its eigenvalues, denoted by $\lambda(\mathcal{L}(\mathcal{G}))$, are all non-negative. Moreover, the null space of $\mathcal{L}(\mathcal{G})$ is spanned by the vector of ones, $\mathbf{1}_n$. Alternatively, the Laplacian can be expressed in terms of the graph's incidence matrix $\mathcal{F}(\mathcal{G})$, such that $\mathcal{L}(\mathcal{G}) = \mathcal{F}(\mathcal{G})\mathcal{F}^\top(\mathcal{G})$.
\begin{lemma}\cite{MesbahiEgerstedt}\label{lem:ac}
	The smallest non-zero eigenvalue of $\mathcal{L}(\mathcal{G})$ (denoted by $\lambda_2(\mathcal{L}(\mathcal{G}))$ and referred to as the algebraic connectivity of the graph, $\mathcal{G}$ satisfies $x^\top\mathcal{L}(\mathcal{G})x\geq \lambda_2(\mathcal{L}(\mathcal{G}))\|x\|_2^2>0$ for some $x\in\mathbb{R}^n$.
\end{lemma}
\begin{definition}
	A switched dynamic network is defined as an ordered pair,  ${\mathcal{G}}_{\eta(t)}=({\mathcal{G}},\eta(t))$ where ${\mathcal{G}}=\{\mathcal{G}_1,\mathcal{G}_2,\ldots,\mathcal{G}_p\}$ is a finite set of graphs sharing a common vertex set and $\eta(t):[t_0,\infty]\to[1,2,\ldots,p]$ is a piecewise continuous switching signal that specifies the active communication topology among the interceptors at each point in time.
\end{definition}
We now briefly discuss the notions of predefined-time stability. Consider a system whose dynamics is given by
\begin{equation}\label{eq:toysystem}
	\dot{z} = -\dfrac{1}{T_s}f(z),~\forall\,t\geq t_0,~f(0)=0,~z(t_0)=z_0,
\end{equation}
where $z\in\mathbb{R}^n$ is the state variable, $T_s>0$ is a parameter and $f:\mathbb{R}^n\to\mathbb{R}^n$ is nonlinear and continuous on $z$ everywhere, except at the origin.
\begin{assumption}\label{asm:psi}
	Let $\Psi(\ell) = \phi(|\ell|)^{-1}\sign(\ell)$ such that $\ell\in\mathbb{R}$ and the function $\phi:\mathbb{R}_+\to{\mathbb{R}}\setminus\{0\}$ satisfies $\phi(0)=\infty$, $\phi(\ell)<\infty\;\forall\,\ell\in\mathbb{R}_+\setminus\{0\}$, and $\int_{0}^{\infty}\phi(\ell)d\ell=1$.
\end{assumption}
\begin{lemma}\cite{ALDANALOPEZ202011880}\label{lem:predeftime}
	If there exists a continuous positive definite radially unbounded function, $\mathscr{V}:\mathbb{R}^n\to\mathbb{R}$, whose time derivative along the trajectories of \eqref{eq:toysystem} satisfies $\dot{\mathscr{V}}(z)\leq-\dfrac{1}{T_s}\Psi(\mathscr{V}(z))$ for all $z\in\mathbb{R}^n\setminus\{0\}$, and $\Psi(z)$ satisfies \Cref{asm:psi}, then the system \eqref{eq:toysystem} is fixed-time stable with a predefined upper bound on the settling time, $T_s$.
\end{lemma}
\begin{lemma}\cite{ALDANALOPEZ202011880}\label{lem:omega}
	A predefined-time consensus function $\Omega:\mathbb{R}\to\mathbb{R}$ is a monotonically increasing function, if there exists a function $\hat{\Omega}:\mathbb{R}_+\to\mathbb{R}_+$, a non-increasing function $\beta:\mathbb{N}\to\mathbb{R}_+$, and a parameter $d\geq 1$, such that for all $z=(z_1,z_2,\ldots,z_n)^\top,~z_i\in\mathbb{R}_+$, the inequality
	\begin{equation}
		\hat{\Omega}\left(\beta(n)\|z\|\right) \leq \beta(n)^d\sum_{i=1}^{n}\Omega(z_i)
	\end{equation}
	holds, and $\Psi(\ell) = \ell^{-1}\hat{\Omega}(|\ell|)$ satisfies \Cref{asm:psi}.
\end{lemma}
\begin{lemma}\label{lem:pdt2}\cite{https://doi.org/10.1002/rnc.4715}
	For a continuous positive definite radially unbounded function, $\mathscr{V}:\mathbb{R}^n\to\mathbb{R}$, such that $\mathscr{V}(0)=0$ and $\mathscr{V}(z)>0$ for all $z\in\mathbb{R}^n\setminus\{0\}$, if the relation
	\begin{align}
		\dot{\mathscr{V}}(z) \leq& - \dfrac{\Gamma\left(\dfrac{1-k\mathfrak{m}}{\mathfrak{n}-\mathfrak{m}}\right)\Gamma\left(\dfrac{k\mathfrak{n}-1}{\mathfrak{n}-\mathfrak{m}}\right)}{T_s\mathscr{M}^k \Gamma(k)\left(\mathfrak{n}-\mathfrak{m}\right)}\left(\dfrac{\mathscr{M}}{\mathscr{N}}\right)^{\frac{1-k\mathfrak{m}}{\mathfrak{n}-\mathfrak{m}}}\nonumber\\
        &\times\left[\mathscr{M}\mathscr{V}(z)^\mathfrak{m} +\mathscr{N}\mathscr{V}(z)^\mathfrak{n} \right]^{k},
	\end{align}
	where $\Gamma(\cdot)$ is the \emph{gamma} function such that $\Gamma(\ell)=\int_{0}^{\infty}e^{-t}t^{\ell-1}dt$, holds for some $\mathscr{M},\mathscr{N},\mathfrak{m},\mathfrak{n},k>0$ satisfying the constraints $k\mathfrak{m}<1$ and $k\mathfrak{n}>1$, then the origin of the system $\dot{z}=f(z),$ is predefined-time stable with a predefined-time, $T_s$.
\end{lemma}
    \begin{lemma}\label{lem:norm}
	For some $z=[z_1,z_2,\ldots,z_n]^\top \in\mathbb{R}^n$ and $p\in\mathbb{N}$, $\|z\|_p = \sqrt[p]{\sum_{i=1}^{n}|z_i|^p}$, such that $\|z\|_a\leq\|z\|_b$ for any $a\geq b$.
\end{lemma}
\begin{lemma}\cite{https://doi.org/10.1002/rnc.4715}\label{lem:jenson2}
	Let $s=(s_1,s_2,\ldots,s_n) \in\mathbb{R}_+$ be a sequence of positive numbers. For some $\mathscr{M},\mathscr{N},\mathfrak{m},\mathfrak{n},k>0$ satisfying the constraints $k\mathfrak{m}<1$ and $k\mathfrak{n}>1$, 
	\begin{align}
		&\left(\dfrac{1}{n}\sum_{i=1}^{n} s_i\right) \left[\mathscr{M}\left(\dfrac{1}{n}\sum_{i=1}^{n} s_i\right)^\mathfrak{m} + \mathscr{N}\left(\dfrac{1}{n}\sum_{i=1}^{n} s_i\right)^\mathfrak{n}\right]^k \nonumber\\
        &\leq \dfrac{1}{n}\sum_{i=1}^{n} s_i\left(\mathscr{M}s_i^\mathfrak{m}+\mathscr{N}s_i^\mathfrak{n}\right)^k.
	\end{align}
\end{lemma}
\begin{assumption}
    The interceptors exchange information over an undirected network, and the switching signal $\eta(t)$ is exogenously generated such that the Zeno behavior is absent (a finite number of switching occurs in a finite time interval).
\end{assumption}

\section{Cooperative Target Capture Guidance}\label{sec:st}

It is important to note that, at any given moment during the engagement, the remaining time until target capture, commonly referred to as the \emph{time-to-go}, plays a critical role in the design of cooperative guidance strategies. However, obtaining an accurate estimate of the same in 3D settings may not be tractable in general, except under special cases.  Assuming a baseline proportional-navigation guidance principle, an expression of time-to-go against a stationary target considering wide launch envelops, which is given by
\begin{align}
    t_{\mathrm{go}_i}(t) = \dfrac{r_i(t)}{V\M(t)}\left(1+\dfrac{\sin^2\sigma\M(t)}{4N_i-2}\right)+w_i(t);~N_i>2\label{eq:tgo}
\end{align}
for the \ith interceptor, where $w_i(t)$ represents the lumped modeling uncertainties from various sources in estimating the time-to-go. Note that the unknown term $w_i(t)$ is usually disregarded in the design of the cooperative guidance strategy, yet it needs to be accounted for in a robust design. In \cite{doi:10.2514/1.G005180}, $w_i(t)$ is identically zero. For brevity, we drop the arguments when it is clear from the context.
\begin{assumption}\label{asm:w}
    The uncertainty $w_i$ is such that $|w_i|\leq w_{\max}<\infty$ and $|\dot{w}_i|\leq \dot{w}_{\max} <\inf_{t\geq 0} \cos\sigma\M\left(1-\frac{\sin^2\sigma\M}{4N_i-2}\right)<\infty$. Moreover, $w_i$ and $\dot{w}_i$ are continuous functions of time.
\end{assumption}
\Cref{asm:w} ensures that the effect of uncertainty and its rate in the guidance design remains limited and well-behaved whenever closing speed is positive. Violations of these bounds can result in erroneous or unstable estimates of time-to-go. One may notice from the upper bound on $\dot{w}_i$ that such situations may make the cooperative engagement dynamics unstable. To mitigate such risks, practical designs must incorporate robust filtering schemes, a discussion of which is outside the scope of the current work.
\begin{lemma}\label{lem:reldeg}
    For the \ith interceptor, the dynamics of the effective lead angle $\sigma\M$, and that of the time-to-go \eqref{eq:tgo}, both have a relative degree of one with respect to the lateral acceleration components in the pitch and the yaw channels.
\end{lemma}
\begin{proof}
    After differentiating \eqref{eq:sigmaM} and \eqref{eq:tgo} with respect to time and performing simplifications using \eqref{eq:engkinematics}, it readily follows that
    \begin{align}
        \dot{\sigma}\M = \dfrac{V\M\sin\sigma\M}{r_i}+\dfrac{\sin\theta\M\cos\psi\M}{V\M\sin\sigma\M}a\M^z+\dfrac{\sin\psi\M}{V\M\sin\sigma\M}a\M^y,
    \end{align}
    whereas
    \begin{align}
        \dot{t}_{\mathrm{go}_i} =& -\cos\sigma\M\left(1-\dfrac{\sin^2\sigma\M}{4N_i-2}\right)+\dfrac{r_i\sin2\theta\M\cos^2\psi\M}{V\M^2\left(4N_i-2\right)}a\M^z \nonumber\\
        &+ \dfrac{r_i\sin2\psi\M\cos\theta\M}{V\M^2\left(4N_i-2\right)}a\M^y+\dot{w}_i,
    \end{align}
    which clearly indicates the relationships of the said variables with the lateral acceleration components.
\end{proof}
Note that the angles $\theta\M$ and $\psi\M$ at values of $0,\pi/2,\pi$ are isolated and belong to sets of measure zero in the continuous domains of possible angles. Even though the coefficients of the lateral acceleration components vanish on thin sets, these sets are not open, not of positive measure, and hence will not persist over time since angles will change as the vehicles maneuver. Therefore, such isolated points do not pose difficulties during the control design. To this end, we define the error in the common time of simultaneous interception for the \ith interceptor as
\begin{align}
    \xi_i =  t_{\mathrm{go}_i} - \left(T_f - t\right),
\end{align}
where $T_f$ is the common impact time that is not specified before homing. The interceptors implicitly agree upon this value as the engagement proceeds.

Using the results in \Cref{lem:reldeg}, we obtain the dynamics of the error variable $\xi_i$ as
\begin{align}
    \dot{\xi}_i = & 1-\cos\sigma\M\left(1-\dfrac{\sin^2\sigma\M}{4N_i-2}\right)+\dfrac{r_i\sin2\theta\M\cos^2\psi\M}{V\M^2\left(4N_i-2\right)}a\M^z \nonumber\\
        &+ \dfrac{r_i\sin2\psi\M\cos\theta\M}{V\M^2\left(4N_i-2\right)}a\M^y+\dot{w}_i,\nonumber\\
        =&1-\cos\sigma\M\left(1-\dfrac{\sin^2\sigma\M}{4N_i-2}\right)+U_i +\dot{w}_i,\label{eq:xidot}
\end{align}
where $U_i$ is the nonlinear combination of the lateral acceleration components or the effective cooperative guidance command that needs to be designed to stabilize $\xi_i$. In this regard, there are two objectives, namely, $\lim_{t\to T_s}|\xi_j-\xi_i|\to 0$ (consensus in the error variable within a predefined time, and hence in the time-to-go values because $\sum_{j\in\mathcal{N}_i}\left(\xi_j-\xi_i\right) = \sum_{j\in\mathcal{N}_i}\left(t_{\mathrm{go}_j}-t_{\mathrm{go}_i}\right)$ holds), and $\lim_{t\to T_f}\xi_i\to 0$ (nullification of time-to-go value at the time of simultaneous target capture), where $T_f>T_s$ for interceptors $i,j$ such that $i\neq j$. Therefore, we now endeavor to design $U_i$, which is presented in the next theorem.
\begin{theorem}
    Consider the cooperative simultaneous target capture scenario described using \eqref{eq:engkinematics}, where the time-to-go is expressed using \eqref{eq:tgo}, and the consensus error dynamics is given by \eqref{eq:xidot}. The effective cooperative guidance command for the \ith interceptor over a switched dynamic network $\mathcal{G}_{\eta(t)}$,
    \begin{align}
        U_i=&-\mathscr{P}_i\sum_{j\in\mathcal{N}_i(\mathcal{G}_{\eta(t)})}\left[\left(\mathscr{M}\left|\left(\xi_j-\xi_i\right)\right|^\mathfrak{m}+\mathscr{N}\left|\left(\xi_j-\xi_i\right)\right|^\mathfrak{n}\right)^k + \mu\right]\nonumber\\
        &\times\sign\left(\xi_j-\xi_i\right)-1+\cos\sigma\M\left(1-\dfrac{\sin^2\sigma\M}{4N_i-2}\right),\label{eq:U}
    \end{align}
    where $\mathscr{P}_i$ satisfies the constraint
    \begin{equation}\label{eq:PiconditionTs}
		\mathscr{P}_i\geq \dfrac{\Gamma\left(\dfrac{1-k\mathfrak{m}}{\mathfrak{n}-\mathfrak{m}}\right)\Gamma\left(\dfrac{k\mathfrak{n}-1}{\mathfrak{n}-\mathfrak{m}}\right)}{T_s\mathscr{M}^k \Gamma(k)\left(\mathfrak{n}-\mathfrak{m}\right)}\left(\dfrac{\mathscr{M}}{\mathscr{N}}\right)^{\frac{1-k\mathfrak{m}}{\mathfrak{n}-\mathfrak{m}}}\left(\dfrac{\munderbar{E}}{\munderbar{\lambda}_2}\right),
	\end{equation}
    such that $\munderbar{E}=\argmin_{b}\left|\mathcal{E}(\mathcal{G}_b)\right|,\,\munderbar{\lambda}_2 = \argmin_{b}\lambda_2(\mathcal{L}(\mathcal{G}_b))$,
	and 
	\begin{equation}\label{eq:mu}
		\mu > \dfrac{\dot{w}_{\max}}{\argmin_{i}\mathscr{P}_i\sqrt{\munderbar{\lambda}}_2},
	\end{equation}
	with $\mathscr{M},\mathscr{N},\mathfrak{m},\mathfrak{n},k,\Gamma(\cdot)$ are defined in \Cref{lem:pdt2}, ensures that the interceptors agree on a common time-to-go value within a predefined time $T_s$ (specified prior to the engagement as a design parameter in the cooperative guidance command) to simultaneously capture a stationary target at a common time $T_f$ (decided via implicit cooperation among interceptors) regardless of the engagement geometry and the lumped bounded uncertainties in the estimation of the time-to-go.
\end{theorem}
\begin{proof}
    On substituting the proposed cooperative guidance command \eqref{eq:U} in the error dynamics \eqref{eq:xidot}, one may obtain
    \begin{align}
        \dot{\xi}_i =& -\mathscr{P}_i\sum_{j\in\mathcal{N}_i(\mathcal{G}_{\eta(t)})}\left[\left(\mathscr{M}\left|\left(\xi_j-\xi_i\right)\right|^\mathfrak{m}+\mathscr{N}\left|\left(\xi_j-\xi_i\right)\right|^\mathfrak{n}\right)^k + \mu\right]\nonumber\\
        &\times\sign\left(\xi_j-\xi_i\right) + \dot{w}_i. \label{eq:xidotwithU}
    \end{align}
    Let the switching signal $\eta(t)=b$ for some $b\in[0,T_s]$, and define a vector $\mathfrak{f}=\mathcal{F}^\top(\mathcal{G}_b)\mathbf{\xi} = [\mathfrak{f}_1,\mathfrak{f}_2,\ldots,\mathfrak{f}_E]^\top$ , where $E$ is the number of edges. Then, one may write
    \begin{equation}
		\varrho(\mathfrak{f}) = \begin{pmatrix}
			-\mathscr{P}_1 \left[\left(\mathscr{M}\left|\mathfrak{f}_1\right|^\mathfrak{m}+\mathscr{N}\left|\mathfrak{f}_1\right|^\mathfrak{n}\right)^k + \mu\right]\sign\left(\mathfrak{f}_1\right) \\
			-\mathscr{P}_2 \left[\left(\mathscr{M}\left|\mathfrak{f}_2\right|^\mathfrak{m}+\mathscr{N}\left|\mathfrak{f}_2\right|^\mathfrak{n}\right)^k + \mu\right]\sign\left(\mathfrak{f}_2\right) \\
			\vdots \\
			-\mathscr{P}_E \left[\left(\mathscr{M}\left|\mathfrak{f}_E\right|^\mathfrak{m}+\mathscr{N}\left|\mathfrak{f}_E\right|^\mathfrak{n}\right)^k + \mu\right]\sign\left(\mathfrak{f}_E\right)
		\end{pmatrix},
	\end{equation}
    while also being able to express \eqref{eq:xidotwithU} in terms of incidence matrix as
    	\begin{equation}\label{eq:xiE}
		\dot{\mathbf{\xi}} = \mathcal{F}(\mathcal{G}_k)\varrho\left(\mathcal{F}^\top(\mathcal{G}_k)\mathbf{\xi}\right)+ \dot{\mathbf{w}},
	\end{equation}
    where $\mathbf{\xi}=[\xi_1,\xi_2,\ldots,\xi_n]^\top$ and $\dot{\mathbf{w}}=[\dot{w}_1,\dot{w}_2,\ldots,\dot{w}_n]^\top$. Note that in \eqref{eq:xiE} and in the steps thereafter, we adopt vector/matrix notation to streamline the proof.

    Now, consider a Lyapunov function candidate $\mathscr{V} (\mathbf{\xi})= \dfrac{\sqrt{\munderbar{\lambda}_2}}{\munderbar{E}}\sqrt{\mathbf{\xi}^\top\mathbf{\xi}}$ such that $\mathscr{V}(0)\iff \mathbf{\xi}\in\mathscr{C}(\mathbf{\xi})$, where the agreement subspace is defined as $\mathscr{C}(\mathbf{\xi})=\{\mathbf{\xi}:\xi_1=\xi_2=\cdots=\xi_n\}$.

    To rigorously establish that the proposed command $U_i$ ensures simultaneous target interception by enforcing a predefined time consensus in the interceptor's time-to-go values, under switched dynamic interactions, it is required to demonstrate that $\mathscr{V}$ constitutes a common Lyapunov function corresponding to all topologies in the set $\mathcal{G}_{\eta(t)}$. On differentiating $\mathscr{V}$ with respect to time, one may obtain
    	\begin{align}
		\dot{\mathscr{V}} =& \dfrac{\sqrt{\munderbar{\lambda}_2}}{\munderbar{E}\sqrt{\mathbf{\xi}^\top\mathbf{\xi}}}\mathbf{\xi}^\top\dot{\mathbf{\xi}} = \dfrac{\sqrt{\munderbar{\lambda}_2}}{\munderbar{E}\sqrt{\mathbf{\xi}^\top\mathbf{\xi}}}\mathbf{\xi}^\top\left[\mathcal{F}(\mathcal{G}_k)\varrho\left(\mathcal{F}^\top(\mathcal{G}_k)\mathbf{\xi}\right) + \dot{\mathbf{w}}\right],\nonumber\\
		=& \dfrac{\sqrt{\munderbar{\lambda}_2}}{\munderbar{E}\|\mathbf{\xi}\|_2}\left[\mathbf{\xi}^\top\mathcal{F}(\mathcal{G}_k)\varrho\left(\mathcal{F}^\top(\mathcal{G}_k)\mathbf{\xi}\right) +\mathbf{\xi}^\top \dot{\mathbf{w}}\right],\nonumber\\
		=&\dfrac{\sqrt{\munderbar{\lambda}_2}}{\munderbar{E}\|\mathbf{\xi}\|_2}\left[\mathfrak{f}^\top\varrho\left(\mathfrak{f}\right) +\mathbf{\xi}^\top \dot{\mathbf{w}}\right],
	\end{align}
    which can be further simplified to
    	\begin{align}
		\dot{\mathscr{V}} =&\dfrac{\sqrt{\munderbar{\lambda}_2}}{\munderbar{E}\|\mathbf{\xi}\|_2}\left[-\sum_{i=1}^{E}\mathscr{P}_i|\mathfrak{f}_i|\left(\mathscr{M}\left|\mathfrak{f}_i\right|^\mathfrak{m}+\mathscr{N}\left|\mathfrak{f}_i\right|^\mathfrak{n}\right)^k \right.\nonumber\\
        &\left.- \mu\sum_{i=1}^{E}\mathscr{P}_i|\mathfrak{f}_i| -\mathbf{\xi}^\top\dot{\mathbf{w}}\right],\nonumber\\
		\leq&\dfrac{\sqrt{\munderbar{\lambda}_2}}{\munderbar{E}\|\mathbf{\xi}\|_2}\left[-\sum_{i=1}^{E}\mathscr{P}_i|\mathfrak{f}_i|\left(\mathscr{M}\left|\mathfrak{f}_i\right|^\mathfrak{m}+\mathscr{N}\left|\mathfrak{f}_i\right|^\mathfrak{n}\right)^k \right.\nonumber\\
        &\left.- \mu\sum_{i=1}^{E}\mathscr{P}_i|\mathfrak{f}_i| +\mathbf{\xi}^\top\dot{w}_{\max}\right].\label{eq:V3dotstep1}
	\end{align}
    Upon letting $\mathscr{P}=\argmin_{i}\mathscr{P}_i$, \eqref{eq:V3dotstep1} can be further written, using the results in \Cref{lem:jenson2}, as
    \begin{align}
		\dot{\mathscr{V}} \leq& \dfrac{\sqrt{\munderbar{\lambda}_2}}{\munderbar{E}\|\mathbf{\xi}\|_2}\left[-\sum_{i=1}^{E}\mathscr{P}|\mathfrak{f}_i|\left(\mathscr{M}\left|\mathfrak{f}_i\right|^\mathfrak{m}+\mathscr{N}\left|\mathfrak{f}_i\right|^\mathfrak{n}\right)^k \right.\nonumber\\
        &\left.- \mu\sum_{i=1}^{E}\mathscr{P}|\mathfrak{f}_i| +\mathbf{\xi}^\top\dot{w}_{\max}\right],\nonumber\\
		\leq& \dfrac{\sqrt{\munderbar{\lambda}_2}}{\munderbar{E}\|\mathbf{\xi}\|_2}\left\{-E\mathscr{P}\left(\dfrac{1}{E}	\sum_{i=1}^{E}|\mathfrak{f}_i|\right)\left[\mathscr{M}\left(\dfrac{1}{E}	\sum_{i=1}^{E}|\mathfrak{f}_i|\right)^\mathfrak{m}\right.\right.\nonumber\\
        &\left.\left.+\mathscr{N}\left(\dfrac{1}{E}	\sum_{i=1}^{E}|\mathfrak{f}_i|\right)^\mathfrak{n}\right]^k- \mu\mathscr{P}\sum_{i=1}^{E}|\mathfrak{f}_i| +\mathbf{\xi}^\top\dot{w}_{\max}\right\}.\label{eq:Vdotstep2}
	\end{align}
    It follows from \Cref{lem:ac,lem:norm} that \eqref{eq:Vdotstep2} may be simplified to
    	\begin{align}
		\dot{\mathscr{V}} \leq& \dfrac{\sqrt{\munderbar{\lambda}_2}}{\munderbar{E}}\left\{-\dfrac{E\mathscr{P}}{\|\mathbf{\xi}\|_2}\left(\dfrac{1}{E}	\|\mathbf{\xi}\|_1\right)\left[\mathscr{M}\left(\dfrac{1}{E}	\|\mathbf{\xi}\|_1\right)^\mathfrak{m}+\mathscr{N}\left(\dfrac{1}{E}	\|\mathbf{\xi}\|_1\right)^\mathfrak{n}\right]^k \right.\nonumber\\ &\left.- \dfrac{\mu\mathscr{P}}{\|\mathbf{\xi}\|_2}\|\mathbf{\xi}\|_1+\dot{w}_{\max}\right\},\nonumber \\
		\leq& \dfrac{\sqrt{\munderbar{\lambda}_2}}{\munderbar{E}}\left\{-\dfrac{E\mathscr{P}}{\|\mathbf{\xi}\|_2}\left(\dfrac{\sqrt{\munderbar{\lambda}}_2}{E}	\|\mathbf{\xi}\|_2\right)\left[\mathscr{M}\left(\dfrac{\sqrt{\munderbar{\lambda}}_2}{E}	\|\mathbf{\xi}\|_2\right)^\mathfrak{m}\right. \right.\nonumber\\
		&\left.\left.+\mathscr{N}\left(\dfrac{\sqrt{\munderbar{\lambda}}_2}{E}	\|\mathbf{\xi}\|_2\right)^\mathfrak{n}\right]^k- \left[\mu\mathscr{P}\sqrt{\munderbar{\lambda}}_2-\dot{w}_{\max}\right] \right\},\nonumber\\
		\leq& \dfrac{\sqrt{\munderbar{\lambda}_2}}{\munderbar{E}}\left\{-\mathscr{P}\sqrt{\munderbar{\lambda}}_2\left[\mathscr{M}\left(\dfrac{\munderbar{\lambda}_2}{\munderbar{E}}	\|\mathbf{\xi}\|_2\right)^\mathfrak{m}+\mathscr{N}\left(\dfrac{\munderbar{\lambda}_2}{\munderbar{E}}	\|\mathbf{\xi}\|_2\right)^\mathfrak{n}\right]^k \right.\nonumber\\
		&\left.- \left[\mu\mathscr{P}\sqrt{\munderbar{\lambda}}_2-\dot{w}_{\max}\right] \right\}.\label{eq:V3dotstep3}
	\end{align}
    If the sufficient condition on $\mu$ given in \eqref{eq:mu} holds, then \eqref{eq:V3dotstep3} becomes
	\begin{align}
		\dot{\mathscr{V}} \leq& -\dfrac{\mathscr{P}{\munderbar{\lambda}_2}}{\munderbar{E}}\left[\mathscr{M}\left(\dfrac{\munderbar{\lambda}_2}{\munderbar{E}}	\|\mathbf{\xi}\|_2\right)^\mathfrak{m}+\mathscr{N}\left(\dfrac{\munderbar{\lambda}_2}{\munderbar{E}}	\|\mathbf{\xi}\|_2\right)^\mathfrak{n}\right]^k,\nonumber\\
        \leq &  -\dfrac{\mathscr{P}{\munderbar{\lambda}_2}}{\munderbar{E}}\left[\mathscr{M}\mathscr{V}^\mathfrak{m}+\mathscr{N}\mathscr{V}^\mathfrak{n}\right]^k,\nonumber\\
		\leq&- \dfrac{\Gamma\left(\dfrac{1-k\mathfrak{m}}{\mathfrak{n}-\mathfrak{m}}\right)\Gamma\left(\dfrac{k\mathfrak{n}-1}{\mathfrak{n}-\mathfrak{m}}\right)}{T_s\mathscr{M}^k \Gamma(k)\left(\mathfrak{n}-\mathfrak{m}\right)}\left(\dfrac{\mathscr{M}}{\mathscr{N}}\right)^{\frac{1-k\mathfrak{m}}{\mathfrak{n}-\mathfrak{m}}}\left[\mathscr{M}\mathscr{V}^\mathfrak{m}+\mathscr{N}\mathscr{V}^\mathfrak{n}\right]^k
	\end{align}
	whenever \eqref{eq:PiconditionTs} is satisfied. The preceding result implies that the interceptors reach consensus in the variables $\xi_i$, and consequently in their time-to-go values, within a predefined time $T_s$ (refer \Cref{lem:pdt2}), irrespective of the underlying interaction topology, as the result holds uniformly for all $\mathcal{G}_b\in\mathcal{G}_{\eta(t)}$. 
    
    Moreover, as $\lim_{t\to T_s}|\xi_j-\xi_i|\to 0$, $\dot{t}_{\mathrm{go}_i}\to -\cos\sigma\M\left(1-\frac{\sin^2\sigma\M}{4N_i-2}\right)+\dot{w}_i.$ To guarantee target interception, $\dot{t}_{\mathrm{go}_i}<0$ eventually. Even if $\dot{w}_i$ is persistently bounded by a constant $\delta \leq \dot{w}_{\max}$, if the rate of the uncertainty satisfies the bound in \Cref{asm:w}, the time-to-go profile for the \ith interceptor will be monotonically decreasing after consensus has been achieved provided $|\sigma\M|\in(-\pi/2,\pi/2)$ after consensus (or equivalently, maintaining a positive closing speed, that is, $\dot{r}_i<0$). It then follows that, using a proper choice of design parameters, the time-to-go profiles can be made to decrease to zero after consensus, eventually guaranteeing a simultaneous interception at a time decided implicitly during the engagement. 
\end{proof}
Recall that $U_i$ is a combination of the lateral acceleration components. This allows the designer a plethora of ways to choose the individual components. In this work, we determine the components $a\M^z,a\M^y$ by solving an optimization problem that minimizes a cost function $\mathcal{J}_i = \left\|\left[\frac{a\M^z}{c_i^z}\;\frac{a\M^y}{c_i^y}\right]^\top\right\|_\ell$,
subject to the constraint $C_i:~\frac{r_i\sin2\theta\M\cos^2\psi\M}{V\M^2\left(4N_i-2\right)}a\M^z+ \frac{r_i\sin2\psi\M\cos\theta\M}{V\M^2\left(4N_i-2\right)}a\M^y=U_i,$ where $c_i^z(t),c_i^y(t)>0$ are time-varying (or constant) positive weights used to regulate the relative control effort in the pitch and yaw channels, and $\|\cdot\|_\ell$ denotes the $\ell$\textsuperscript{th} norm in $\mathbb{R}^2$ at a given time instant. Such a cost function models trade-offs between control authority, energy efficiency, and actuator prioritization in multi-input systems, whereas the choice of a specific $\ell$ governs how the aggregate control effort is quantified. We present the explicit solutions of $a\M^z,a\M^y$ for specific values of $\ell$ next.
\begin{theorem}\label{thm:amL}
    The lateral acceleration components for the \ith interceptor obtained after minimization of $\mathcal{J}$ for $\ell\in(1,\infty)$ subject to the affine equality constraint $C_i$ are
    \begin{align}
		a\M^z =& \dfrac{V\M^2\left(4N_i-2\right)\left\vert\sin2\theta\M\cos^2\psi\M\right\vert^{\frac{1}{\ell-1}}\sign(\sin2\theta\M)}{r_i\left(\left\vert\sin 2\theta\M\cos^2\psi\M\right\vert^{\frac{\ell}{\ell-1}}+\left\vert c_i\sin2\psi\M\cos\theta\M\right\vert^{\frac{\ell}{\ell-1}}\right)}\nonumber\\
        &\left\{-\mathscr{P}_i\sum_{j\in\mathcal{N}_i(\mathcal{G}_{\eta(t)})} \left[\left(\mathscr{M}\left|\left(\xi_j-\xi_i\right)\right|^\mathfrak{m}+\mathscr{N}\left|\left(\xi_j-\xi_i\right)\right|^\mathfrak{n}\right)^k \right.\right.\nonumber\\
        &\left.\left.+ \mu\right]\sign\left(\xi_j-\xi_i\right)-1+\cos\sigma\M\left(1-\dfrac{\sin^2\sigma\M}{4N_i-2}\right)\right\},\label{eq:amz}\\ 
  a\M^y =& \dfrac{c_i^\ell V\M^2\left(4N_i-2\right)\left\vert\sin2\psi\M\cos\theta\M\right\vert^{\frac{1}{\ell-1}}\sign(\sin2\psi\M\cos\theta\M)}{r_i\left(\left\vert\sin 2\theta\M\cos^2\psi\M\right\vert^{\frac{\ell}{\ell-1}}+\left\vert c_i\sin2\psi\M\cos\theta\M\right\vert^{\frac{\ell}{\ell-1}}\right)}\nonumber\\
        &\left\{-\mathscr{P}_i\sum_{j\in\mathcal{N}_i(\mathcal{G}_{\eta(t)})} \left[\left(\mathscr{M}\left|\left(\xi_j-\xi_i\right)\right|^\mathfrak{m}+\mathscr{N}\left|\left(\xi_j-\xi_i\right)\right|^\mathfrak{n}\right)^k \right.\right.\nonumber\\
        &\left.\left.+ \mu\right]\sign\left(\xi_j-\xi_i\right)-1+\cos\sigma\M\left(1-\dfrac{\sin^2\sigma\M}{4N_i-2}\right)\right\},\label{eq:amy}
	\end{align}
	where $c_i={c_i^y}/{c_i^z}$.
\end{theorem}
\begin{proof}
    Since $U_i=\frac{r_i\sin2\theta\M\cos^2\psi\M}{V\M^2\left(4N_i-2\right)}a\M^z+ \frac{r_i\sin2\psi\M\cos\theta\M}{V\M^2\left(4N_i-2\right)}a\M^y$, it follows that $a\M^y=\dfrac{U_i - b_i^z a\M^z}{b_i^y}$, where $b_i^z=\frac{r_i\sin2\theta\M\cos^2\psi\M}{V\M^2\left(4N_i-2\right)}$ and $b_i^y=\frac{r_i\sin2\psi\M\cos\theta\M}{V\M^2\left(4N_i-2\right)}$ such that $\mathcal{J}=\sqrt[\ell]{\left(a\M^z/c_i^z\right)^\ell + \left((U_i - b_i^z a\M^z)/(b_i^yc_i^y)\right)^\ell}$. Partial differentiation of $\mathcal{J}$ with respect to $a\M^z$ and setting the result to zero leads us to arrive at \eqref{eq:amz} after substituting for $U_i$ from \eqref{eq:U}. Thereafter, $a\M^y$ can be obtained using the above relationship. It can be readily verified by the second derivative test that the values thus obtained are indeed minimum. We omit detailed steps due to space constraints.
\end{proof}
For $\ell=2$, the cost $\mathcal{J}$ reflects the Euclidean norm, yielding a balanced distribution of control effort that minimizes total control energy or power. We now proceed to evaluate the optimal values of $a\M^z,a\M^y$ for special cases $\ell=1,\ell\to \infty$, which require distinct analytical treatment due to the inherent non-differentiability of these norms under conventional calculus. Specifically, the $\ell_1-$norm involves absolute value operations, which are non-smooth and lack classical derivatives at certain points. Likewise, the $\ell_\infty-$norm is defined via a maximum operator, whose gradient is not well-defined at points where the maximum is attained by multiple arguments. Consequently, standard optimization techniques based on gradient information must be adapted or replaced with nonsmooth analysis tools in these cases. We summarize the results for these cases next.
\begin{theorem}\label{thm:amL1}
    The lateral acceleration components for the \ith interceptor obtained after minimization of $\mathcal{J}$ for $\ell=1$ subject to the affine equality constraint $C_i$ are
    \begin{align}
        \begin{bmatrix}
a\M^z &
a\M^y
\end{bmatrix}^\top
=
\begin{cases}
\begin{bmatrix}
0 &
\dfrac{U_i}{b_i^y}
\end{bmatrix}^\top; & \mathrm{if}~ c_i > \left\lvert \dfrac{b_i^z}{b_i^y} \right\rvert \\
\begin{bmatrix}
\dfrac{U_i}{b_i^z} &
0
\end{bmatrix}^\top; & \mathrm{if}~ c_i < \left\lvert \dfrac{b_i^z}{b_i^y} \right\rvert
\end{cases},\label{eq:amL1}
    \end{align}
    where $b_i^z,b_i^y$ are defined in the proof of \Cref{thm:amL}, and $U_i$ is given by \eqref{eq:U}.
\end{theorem}
\begin{proof}
    For $\ell=1$, $\mathcal{J}=\left\vert\dfrac{a\M^z}{c_i^z}\right\vert+\left\vert\dfrac{U_i - b_i^za\M^z}{c_i^yb_i^y}\right\vert$, which is convex, piecewise-linear function of $a\M^z$ representing the sum of two absolute value functions. Hence, any global minimizer lies either at a nondifferentiable point or at a breakpoint. We now evaluate $\mathcal{J}$ at two natural candidates derived from setting $a\M^z=0$ and $a\M^y=0$, which both lie in the feasible set $\left\{\left(a\M^z,a\M^y\right)\in\mathbb{R}^2~\bigg\vert~b_i^za\M^z+b_i^ya\M^y=U_i\right\}$. For $a\M^z=0$, one has $a\M^y=U_i/b_i^y$ and $\mathcal{J}=\left\vert\dfrac{U_i}{c_i^yb_i^y}\right\vert=\mathcal{J}_1$ (say). Similarly, for $a\M^y=0$, we get $a\M^z=U_i/b_i^z$ and $\mathcal{J}=\left\vert\dfrac{U_i}{c_i^zb_i^z}\right\vert=\mathcal{J}_2$ (say). If $\mathcal{J}_1<\mathcal{J}_2$, the minimum value of $a\M^z$ is achieved at $\left(0,U_i/b_i^y\right)$, whereas the minimum value of $a\M^y$ is obtained as $\left(U_i/b_i^z,0\right)$ when $\mathcal{J}_1>\mathcal{J}_2$. Reconciling these separate cases, one obtains the conditions in \eqref{eq:amL1}.
\end{proof}
It is worth noting that for $\ell=1$, the cost $\mathcal{J}$ corresponds to the absolute sum of weighted control magnitudes, wherein the interceptors utilize only a single (the most cost-effective) control at a given instant of time to achieve a simultaneous interception. This type of control allocation essentially promotes minimal total energy usage.
\begin{theorem}\label{thm:amLinfty}
    The lateral acceleration components for the \ith interceptor obtained after minimization of $\mathcal{J}$ for $\ell\to\infty$ subject to the affine equality constraint $C_i$ are
    \begin{align}
        \begin{bmatrix}
a\M^z &
a\M^y
\end{bmatrix}^\top
=
\begin{cases}
\begin{bmatrix}
\dfrac{U_i}{b_i^z-c_i b_i^y} &
\dfrac{-c_iU_i}{b_i^z-c_i b_i^y}
\end{bmatrix}^\top; & \dfrac{b_i^z}{b_i^y}<0  \\
\begin{bmatrix}
\dfrac{U_i}{b_i^z+c_i b_i^y} &
\dfrac{c_iU_i}{b_i^z+c_i b_i^y}
\end{bmatrix}^\top; & \dfrac{b_i^z}{b_i^y} >0
\end{cases},\label{eq:amLinfty}
    \end{align}
    where $b_i^z,b_i^y$ are defined in the proof of \Cref{thm:amL}, and $U_i$ is given by \eqref{eq:U}.
\end{theorem}
\begin{proof}
    For $\ell\to\infty$, $\mathcal{J}=\max\left\{ \left\vert{a\M^z}/{c_i^z}\right\vert,\left\vert{a\M^y}/{c_i^y}\right\vert\right\} =  \max\left\{ \left\vert\dfrac{a\M^z}{c_i^z}\right\vert,\left\vert\dfrac{U_i - b_i^za\M^z}{c_i^yb_i^y}\right\vert\right\}$. Since both arguments are continuous and convex in $a\M^z$, and knowing that the pointwise maximum of convex functions is convex, we conclude $\mathcal{J}$ is convex. Therefore, a global minimum occurs where the two arguments are equal (assuming differentiability or continuity of subgradients), i.e., $\left\vert\dfrac{a\M^z}{c_i^z}\right\vert=\left\vert\dfrac{U_i - b_i^za\M^z}{c_i^yb_i^y}\right\vert$. On solving the preceding expression and grouping terms based on sign leads one to arrive at \eqref{eq:amLinfty}.
\end{proof}
For $\ell\to\infty$, the cost $\mathcal{J}$ approaches the maximum of the weighted control magnitudes, which effectively minimizes the peak actuator demand at a given instant of time. This type of allocation may be desirable in resource-constrained or safety-critical environments and provides robustness to loss of control authority due to saturation. 

To vindicate the nonsingularity of the proposed guidance strategy and to show that the isolated points pose no limitations during implementation, we now proceed with a detailed analysis of the dynamic behavior of the relevant engagement variables in the cooperative guidance commands.
\begin{theorem}
    The lateral acceleration components obtained after minimization of $\mathcal{J}$ are nonsingular, except in sets of measure zero.
\end{theorem}
\begin{proof}
    To analyze the nonsingularity of the lateral acceleration components, we refer to the expression in \Cref{thm:amL} and establish that $\left\vert\sin 2\theta\M\cos^2\psi\M\right\vert^{\frac{\ell}{\ell-1}}+\left\vert c_i\sin2\psi\M\cos\theta\M\right\vert^{\frac{\ell}{\ell-1}}= 0$ if and only if the two terms are zero individually. Suppose, for contradiction, that the expression equals zero. Then both terms must vanish, which implies that  $\sin 2\theta\M\cos^2\psi\M=0$ and $\sin2\psi\M\cos\theta\M=0$. These conditions occur only at isolated values of $\theta\M,\psi\M$, specifically at $0,\pi/2,\pi$. Hence, for all $\theta\M,\psi\M\in\mathbb{R}$ excluding this measure-zero set, at least one of the two terms ($b_i^z$ and $b_i^y$) is strictly positive. Since $\ell>1$ and $c_i>0$, both exponents and coefficients preserve positivity, and we can conclude that $\left\vert\sin 2\theta\M\cos^2\psi\M\right\vert^{\frac{\ell}{\ell-1}}+\left\vert c_i\sin2\psi\M\cos\theta\M\right\vert^{\frac{\ell}{\ell-1}}> 0~\forall~\theta\M,\psi\M\in\mathbb{R}$ except in a discrete set. In particular, the expression is non-zero almost everywhere and strictly positive under typical assumptions. From this argument and the structure of the lateral acceleration components obtained in \Cref{thm:amL1,thm:amLinfty}, it follows that those components are also nonsingular by design.
\end{proof}
One may note that $\sin 2\theta\M\cos^2\psi\M=0$ and $\sin2\psi\M\cos\theta\M=0$ simultaneously may occur when both $\theta\M=0$ and $\psi\M=0$. However, this essentially means that $\sigma\M=0$, as can be seen from \eqref{eq:sigmaM}. We now analyze the dynamics of \eqref{eq:sigmaM} after consensus in time-to-go has been established to gain further insights into the behavior of $\sigma\M$ and rule out the possibility of potential singularities. After consensus in time-to-go (after $t\geq T_s$), $\xi_i=0$. Therefore, after differentiating \eqref{eq:tgo}, one may write
\begin{align}
    \dot{\xi}= 1+\dfrac{\dot{r}_i}{V\M}\left(1+\dfrac{\sin^2\sigma\M}{4N_i-2}\right)+\dfrac{{r}_i}{V\M}\left(\dfrac{\sin2\sigma\M\dot{\sigma}\M}{4N_i-2}\right)+\dot{w}_i=0\nonumber
\end{align}
to obtain the dynamics of the effective lead angle for the \ith interceptor as
\begin{align}
    \dot{\sigma}\M=\dfrac{V\M\left(4N_i-2\right)}{r_i\sin2\sigma\M}\left[-1-\dot{w}_i+\cos\sigma\M\left(1+\dfrac{\sin^2\sigma\M}{4N_i-2}\right)\right],\label{eq:sigmadotss}
\end{align}
which can be shown to be monotonically decreasing to zero in the endgame.
\begin{theorem}\label{thm:sigmadot}
    After consensus in time-to-go has been achieved, the equilibrium point $\sigma\M=0$ is globally asymptotically stable in the open interval $\left(-\pi/2,\pi/2\right)$.
\end{theorem}
\begin{proof}
    We consider a Lyapunov function candidate $\mathcal{W}_i=\sigma\M^2/2$, whose time differentiation yields $\dot{\mathcal{W}}_i=\sigma\M\dot{\sigma}\M$. Consequently, after some trigonometric manipulations and simplifications, one has
    \begin{align}
        \dot{\mathcal{W}}_i\leq&-\left(\dfrac{V\M}{2r_i}\right)\dfrac{\sigma}{2}\sin\dfrac{\sigma\M}{2}\left[\dfrac{4N_i-2}{\cos\dfrac{\sigma\M}{2}\cos\sigma\M}-2\cos\dfrac{\sigma\M}{2}\right]\nonumber\\
        &-\dfrac{V\M\left(4N_i-2\right)\dot{w}_{\max}}{r_i},
    \end{align}
    which is clearly negative for $\sigma\M\neq 0$ (following \Cref{asm:w}) since the term in square brackets is strictly positive. This indicates that $\sigma\M\to 0$ after consensus in time-to-go is established, regardless of its initial sign or magnitude (within the said interval), and any method of control allocation.
\end{proof}
It follows from \Cref{thm:sigmadot} that $\sigma\M\to 0$ and hence $\sin 2\theta\M\cos^2\psi\M$ and $\sin2\psi\M\cos\theta\M$ cannot be simultaneously zero unless $\sigma\M=0$. This can arise in two cases. The first case is when the initial value of $\sigma\M=0$, which is a unique case, wherein the interceptors are already aligned on the LOS to the target, and will follow a straight line trajectory. However, in all other scenarios, even if $\sigma\M$ slightly differs from zero, then it will vary accordingly and decrease after consensus in time-to-go, preventing any singularities.

\section{Simulations}\label{sec:simulations}
We now demonstrate the performance of the proposed approach via simulations. In each case, the target is at the origin of the inertial frame, which is radially separated from each interceptor by 10 km. There are five interceptors labeled $I_1$ through $I_5$, whose speeds are heterogeneous, taken as $V\M \in [400~405~390~385~395]^\top$ m/s. The initial elevation angles of the LOS are $[45~-45~135~-135~0]^\top$ degrees, whereas the initial azimuth angles are $[0~60~-75~-10~-20]^\top$ degrees. The initial heading angles of the interceptors are such that $\theta\M \in [120~-60~10~75~-100]^\top$ degrees, whereas $\psi\M\in [55~15~-45~-30~-60]^\top$ degrees. The weights in each channel are kept unity unless noted otherwise. The controller parameters remain fixed at $\mathscr{M}=1,\mathscr{N}=5,\mathfrak{m}=0.1,\mathfrak{n}=2,k=2$, and the time of consensus, $T_s$, is chosen as $3$ s. The bounds on available lateral acceleration are set to $\pm 40$ g, where g is the acceleration due to gravity. The network is arbitrarily switched within three different topologies shown in \Cref{fig:switchingtopologies} based on $\eta(t) = \{1,2,3\}$. 
\begin{figure}[h!]
		\begin{subfigure}[t]{0.32\linewidth}
			\centering
			\resizebox{\linewidth}{!}{
				\begin{tikzpicture}
					\tikzstyle{vertex} = [circle,draw=black,fill=blue!20]
					\tikzstyle{edge} = [-,>=stealth',shorten >=0pt,  thick, auto]
					\tikzstyle{dashedEdge} = [->,>=stealth',shorten >=0pt, bend left, thick, dotted, auto]
					\tikzstyle{writeBelow} = [sloped, anchor=center, below]
					\tikzstyle{writeAbove} = [sloped, anchor=center, above]
				
					\foreach \phi in {1,...,5}{
						\node[vertex] (v_\phi) at (360/5 * \phi:3cm) {$\phi$};

					}
					
					\draw[edge]
					(v_1) edge node[writeAbove] {} (v_3) 
					(v_1) edge node[writeAbove] {} (v_4)
					(v_2) edge node[writeBelow] {} (v_5)
					(v_3) edge node[writeBelow] {} (v_4)
					(v_3) edge node[writeAbove] {} (v_5)
					;		
				\end{tikzpicture}%
			}
			\caption{Active for $\eta=1$.}
			\label{fig:graph1}
		\end{subfigure}
		\begin{subfigure}[t]{0.32\linewidth}
			\centering
			\resizebox{\linewidth}{!}{
				\begin{tikzpicture}
					\tikzstyle{vertex} = [circle,draw=black,fill=blue!20]
					\tikzstyle{edge} = [-,>=stealth',shorten >=0pt,  thick, auto]
					\tikzstyle{dashedEdge} = [->,>=stealth',shorten >=0pt, bend left, thick, dotted, auto]
					\tikzstyle{writeBelow} = [sloped, anchor=center, below]
					\tikzstyle{writeAbove} = [sloped, anchor=center, above]
					
					\foreach \phi in {1,...,5}{
						\node[vertex] (v_\phi) at (360/5 * \phi:3cm) {$\phi$};

					}
					
					\draw[edge]
					(v_1) edge node[writeAbove] {} (v_2) 
					(v_1) edge node[writeAbove] {} (v_3)
					(v_1) edge node[writeAbove] {} (v_5)
					(v_2) edge node[writeBelow] {} (v_4)
					(v_2) edge node[writeBelow] {} (v_5)
					(v_3) edge node[writeAbove] {} (v_5)
					;		
				\end{tikzpicture}%
			}
			\caption{Active for $\eta=2$.}
			\label{fig:graph2}
		\end{subfigure}
		\begin{subfigure}[t]{0.32\linewidth}
			\centering
			\resizebox{\linewidth}{!}{
				\begin{tikzpicture}
					\tikzstyle{vertex} = [circle,draw=black,fill=blue!20]
					\tikzstyle{edge} = [-,>=stealth',shorten >=0pt, bend left, thick, auto]
					\tikzstyle{dashedEdge} = [->,>=stealth',shorten >=0pt, bend left, thick, dotted, auto]
					\tikzstyle{writeBelow} = [sloped, anchor=center, below]
					\tikzstyle{writeAbove} = [sloped, anchor=center, above]
					
					\foreach \phi in {1,...,5}{
						\node[vertex] (v_\phi) at (360/5 * \phi:3cm) {$\phi$};

					}
					
					\draw[edge]
					(v_2) edge node[writeAbove] {} (v_1) 
					(v_3) edge node[writeAbove] {} (v_2)
					(v_4) edge node[writeBelow] {} (v_3)
					(v_5) edge node[writeBelow] {} (v_4)
					(v_1) edge node[writeAbove] {} (v_5)
					;		
				\end{tikzpicture}%
			}
			\caption{Active for $\eta=3$.}
			\label{fig:cycle3}
		\end{subfigure}
	\caption{Interceptors' switching network topologies.}
	\label{fig:switchingtopologies}
	\end{figure}
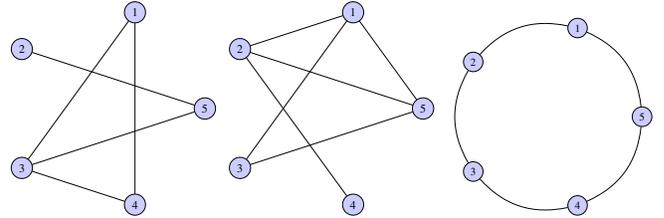
    \begin{figure*}[ht!]
		\begin{subfigure}[t]{0.245\linewidth}
			\centering
			\includegraphics[width=1.1\linewidth]{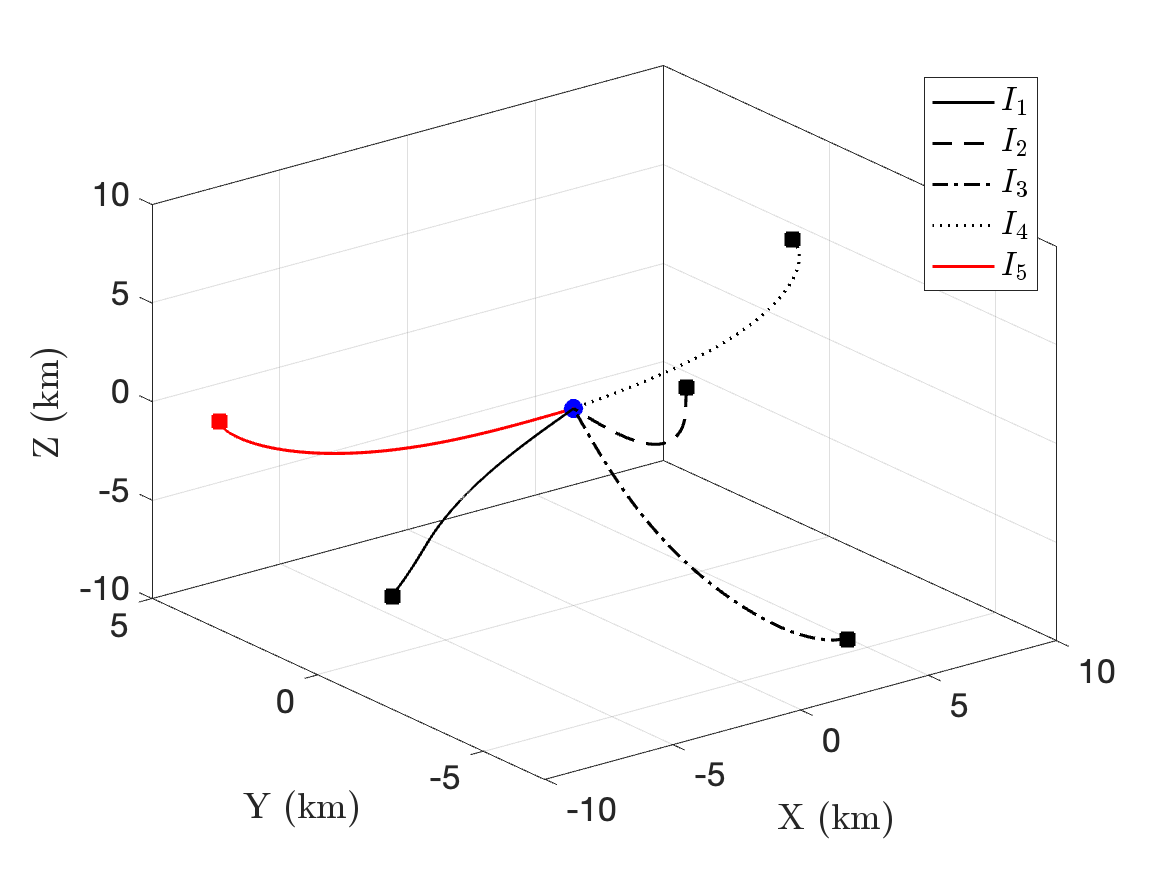}
			\caption{}
			\label{fig:L5trajectory}
		\end{subfigure}
		\hfill
		\begin{subfigure}[t]{0.245\linewidth}
			\centering
			\includegraphics[width=1.1\linewidth]{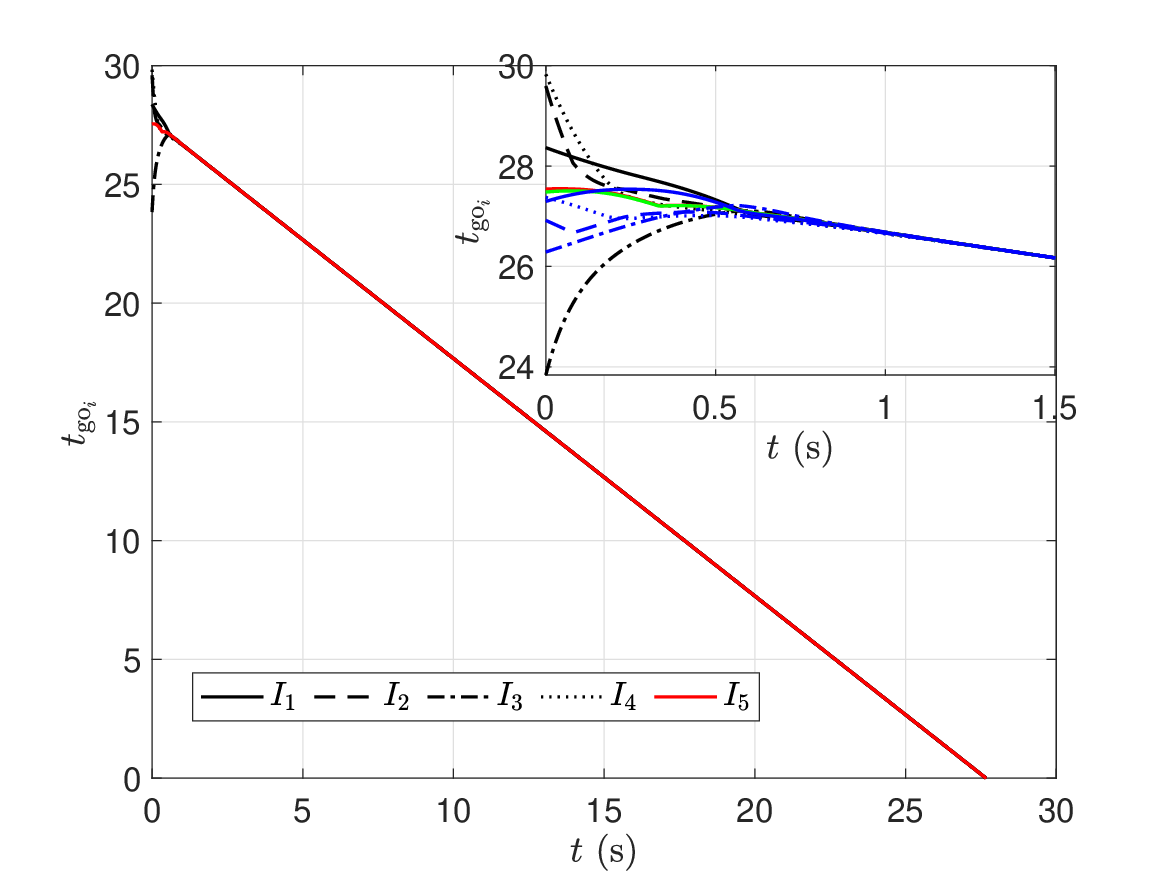}
			\caption{}
			\label{fig:L5tgo}
		\end{subfigure}
		\begin{subfigure}[t]{0.245\linewidth}
			\centering
			\includegraphics[width=1.1\linewidth]{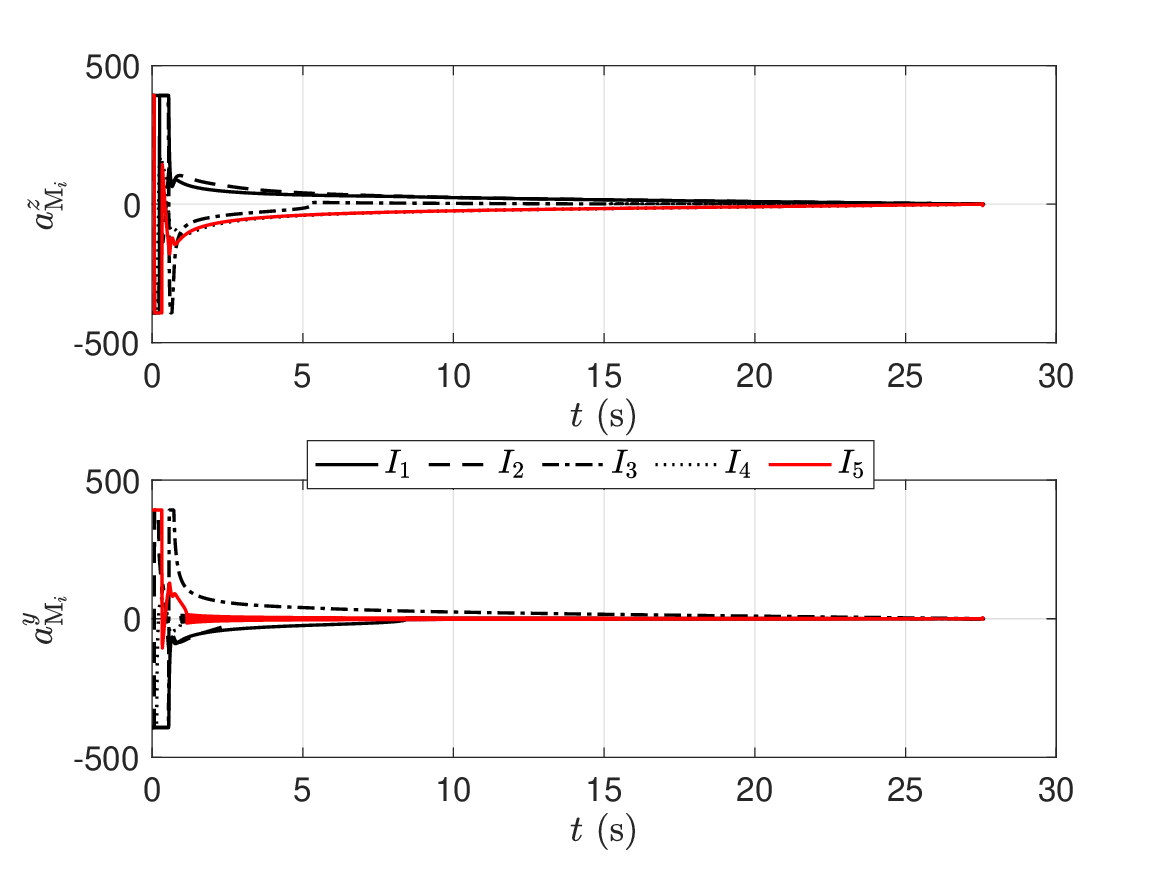}
			\caption{}
			\label{fig:L5am}
		\end{subfigure}
		\hfill
		\begin{subfigure}[t]{0.245\linewidth}
			\centering
			\includegraphics[width=1.1\linewidth]{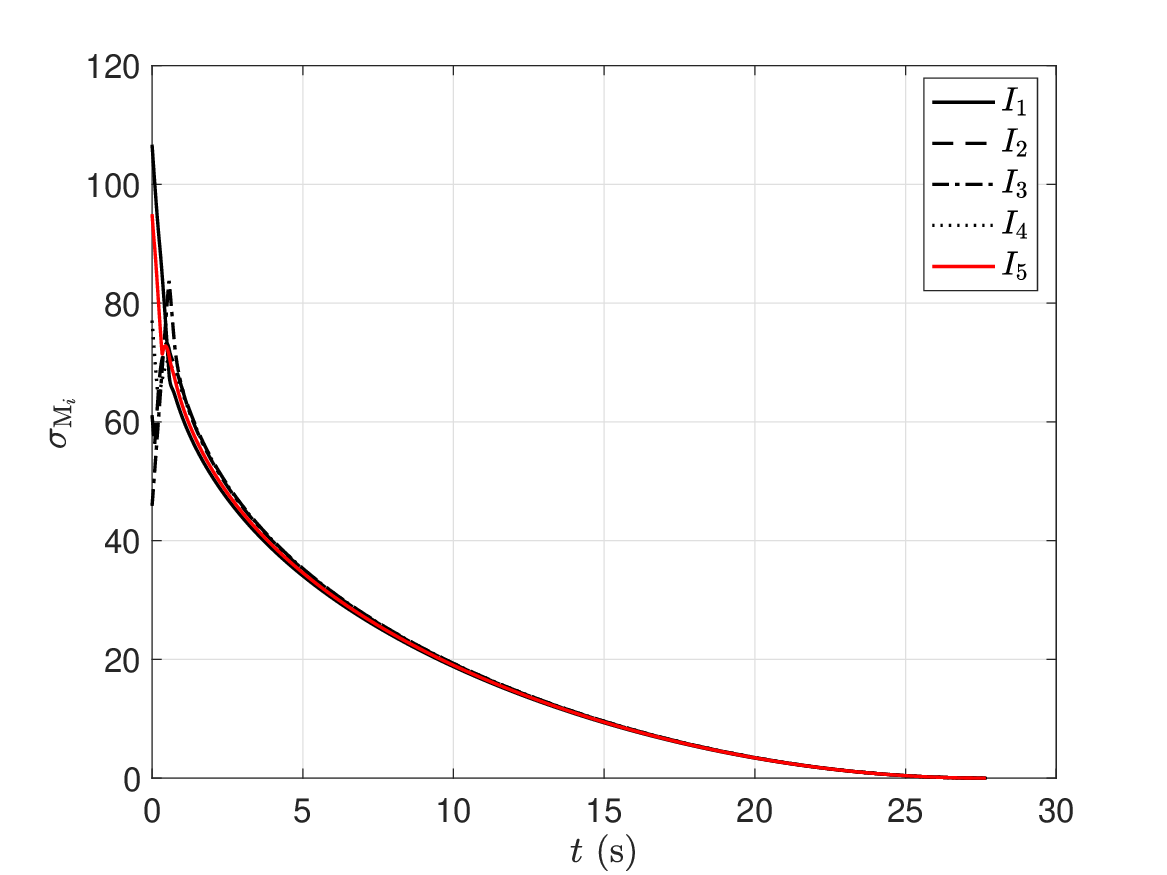}
			\caption{}
			\label{fig:L5sigma}
		\end{subfigure}
		\caption{Cooperative simultaneous target interception (control effort allocation for $\ell=5$ and uncertainty $w_i\neq 0$).}
		\label{fig:L5}
	\end{figure*}
    \begin{figure*}[ht!]
		\begin{subfigure}[t]{0.245\linewidth}
			\centering
			\includegraphics[width=1.1\linewidth]{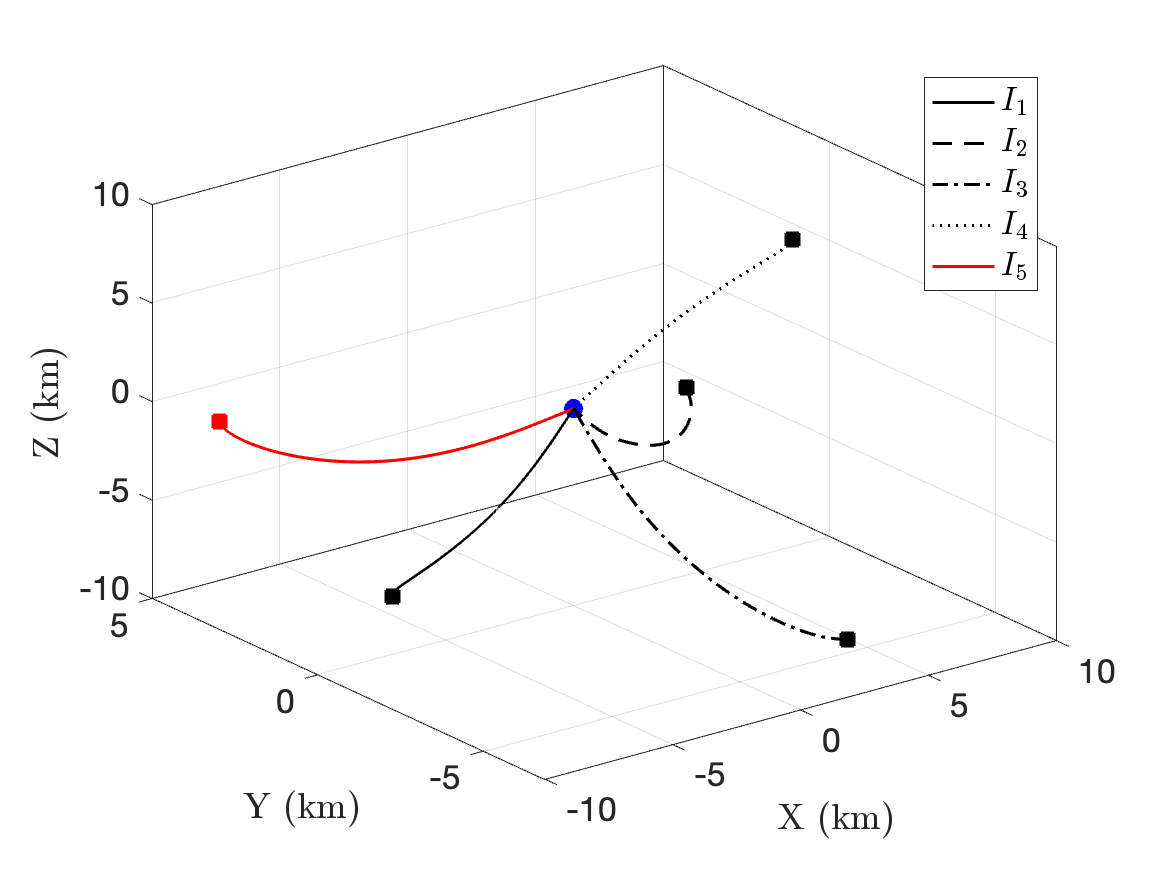}
			\caption{}
			\label{fig:L1trajectory}
		\end{subfigure}
		\hfill
		\begin{subfigure}[t]{0.245\linewidth}
			\centering
			\includegraphics[width=1.1\linewidth]{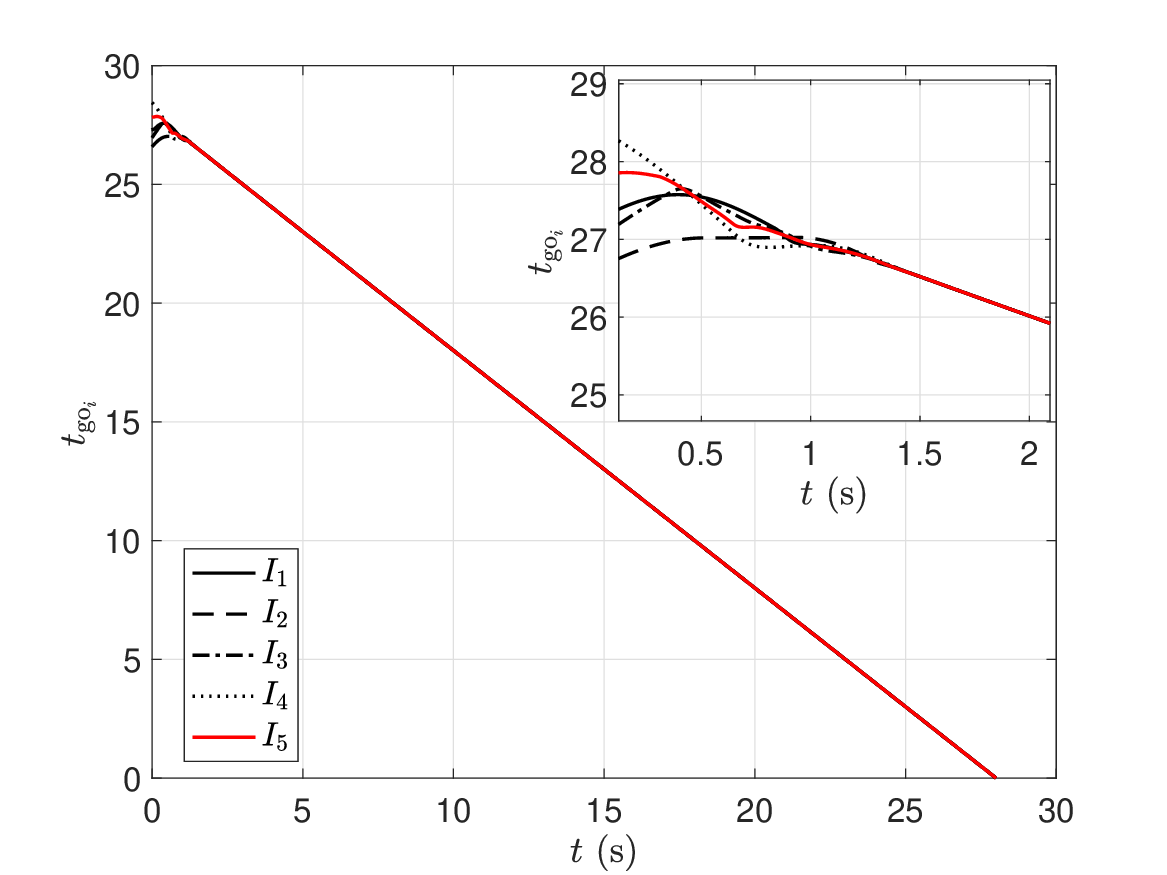}
			\caption{}
			\label{fig:L1tgo}
		\end{subfigure}
		\begin{subfigure}[t]{0.245\linewidth}
			\centering
			\includegraphics[width=1.1\linewidth]{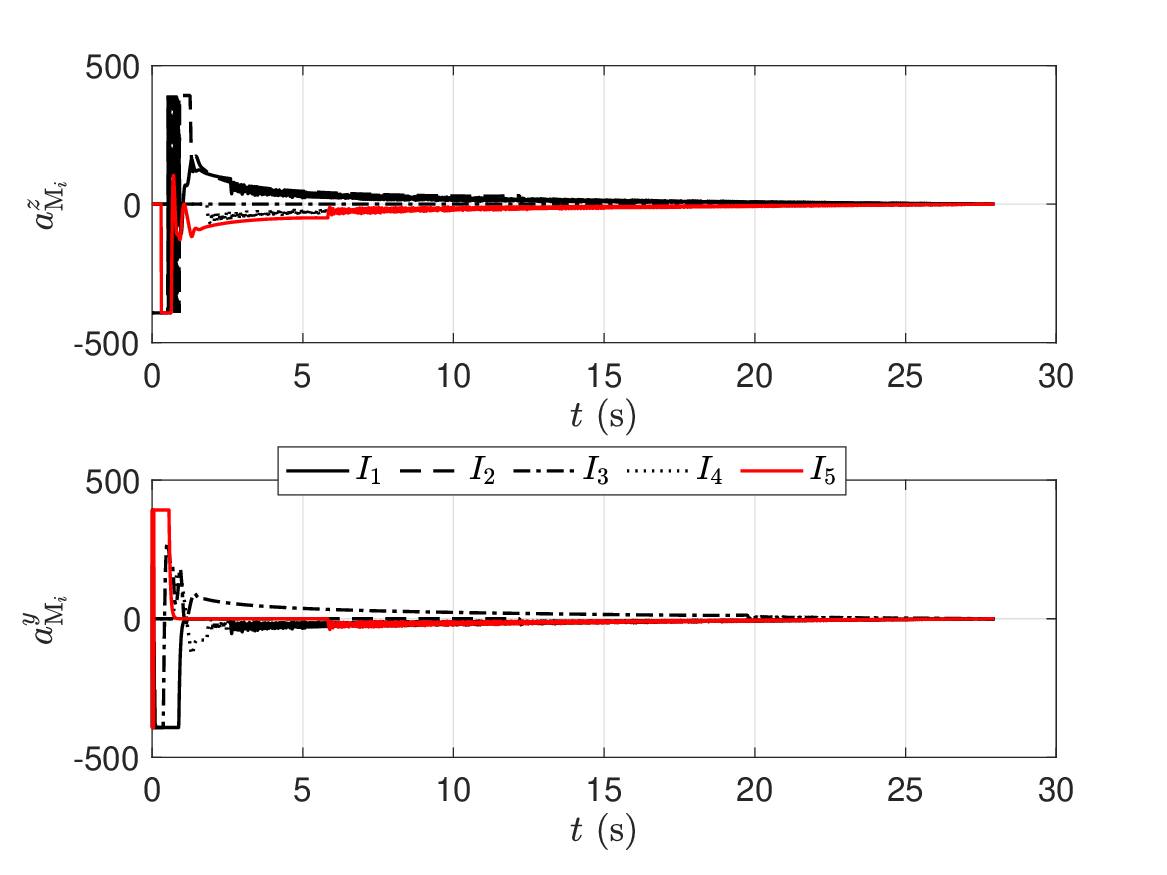}
			\caption{}
			\label{fig:L1am}
		\end{subfigure}
		\hfill
		\begin{subfigure}[t]{0.245\linewidth}
			\centering
			\includegraphics[width=1.1\linewidth]{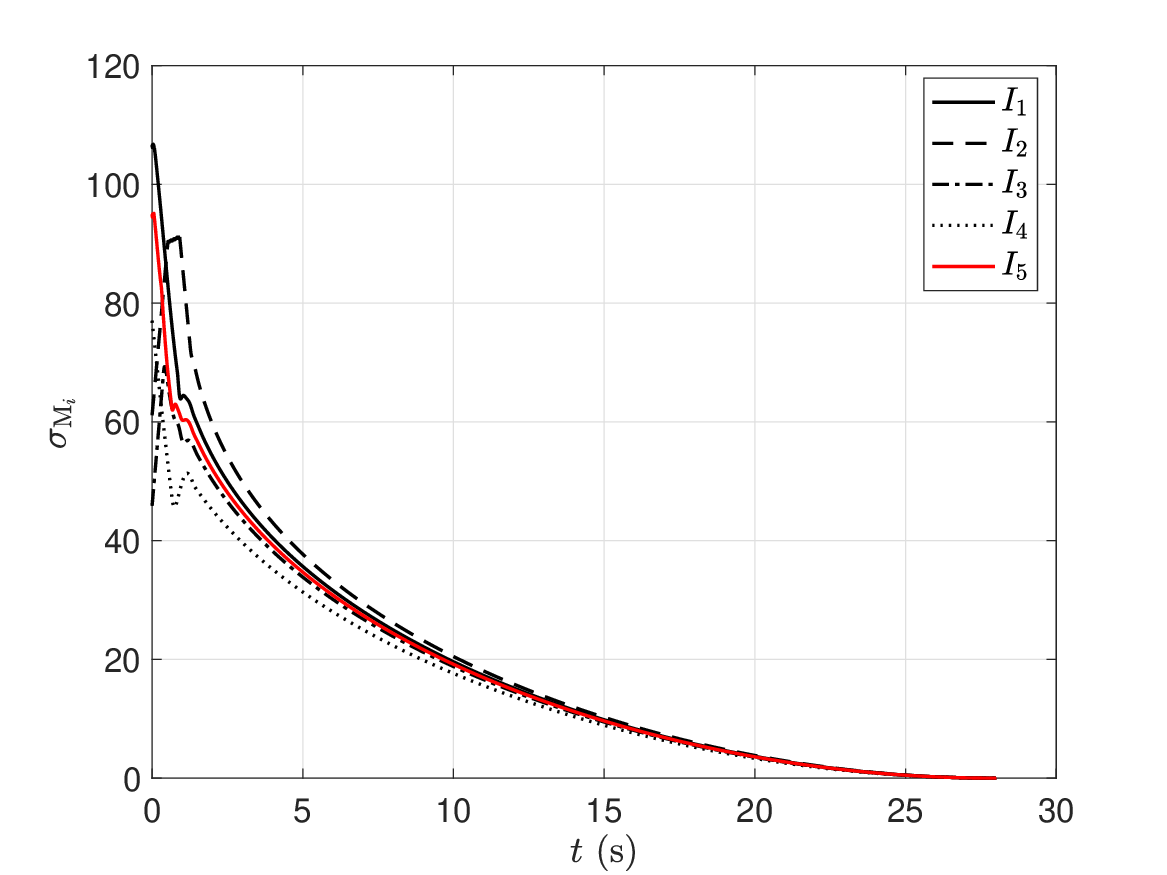}
			\caption{}
			\label{fig:L1sigma}
		\end{subfigure}
		\caption{Cooperative simultaneous target interception (control effort allocation for $\ell=1$ and uncertainty $w_i= 0$).}
		\label{fig:L1}
	\end{figure*}
     \begin{figure*}[ht!]
		\begin{subfigure}[t]{0.245\linewidth}
			\centering
			\includegraphics[width=1.1\linewidth]{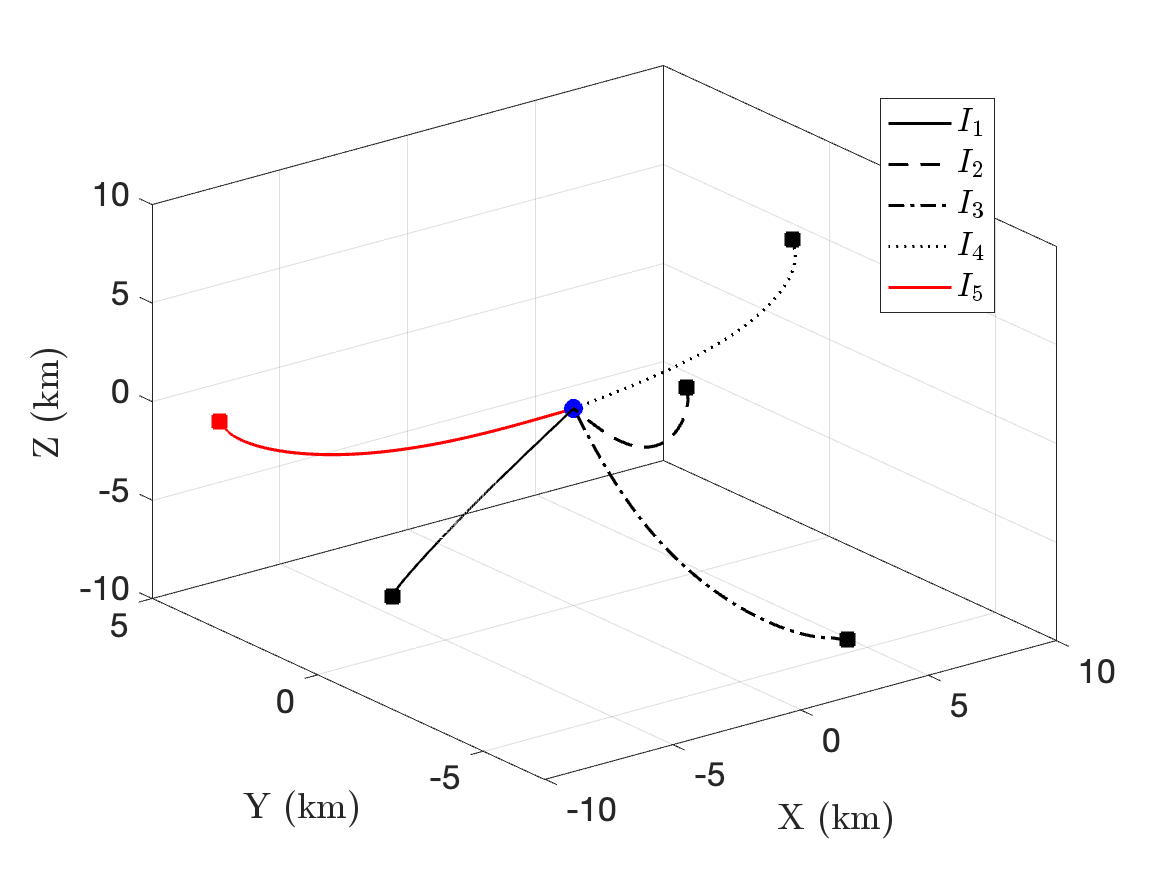}
			\caption{}
			\label{fig:L2trajectory}
		\end{subfigure}
		\hfill
		\begin{subfigure}[t]{0.245\linewidth}
			\centering
			\includegraphics[width=1.1\linewidth]{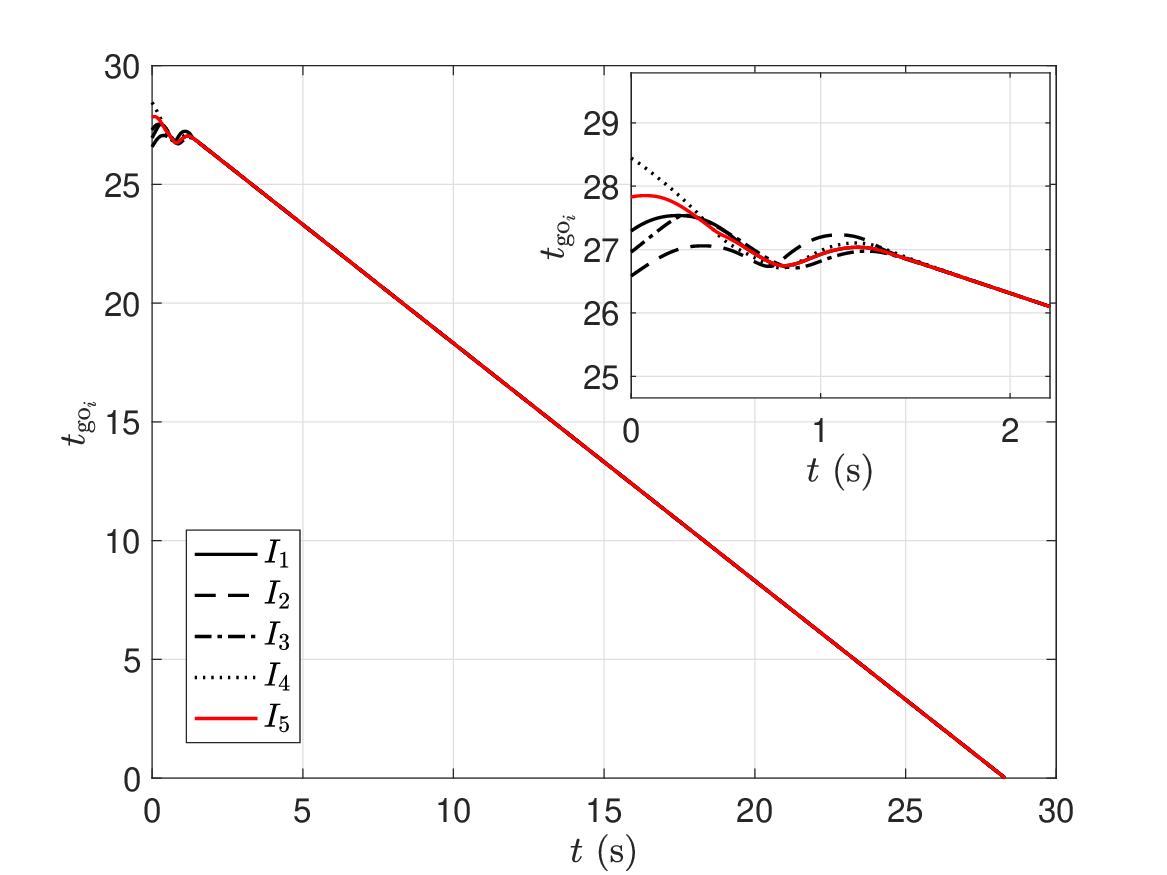}
			\caption{}
			\label{fig:L2tgo}
		\end{subfigure}
		\begin{subfigure}[t]{0.245\linewidth}
			\centering
			\includegraphics[width=1.1\linewidth]{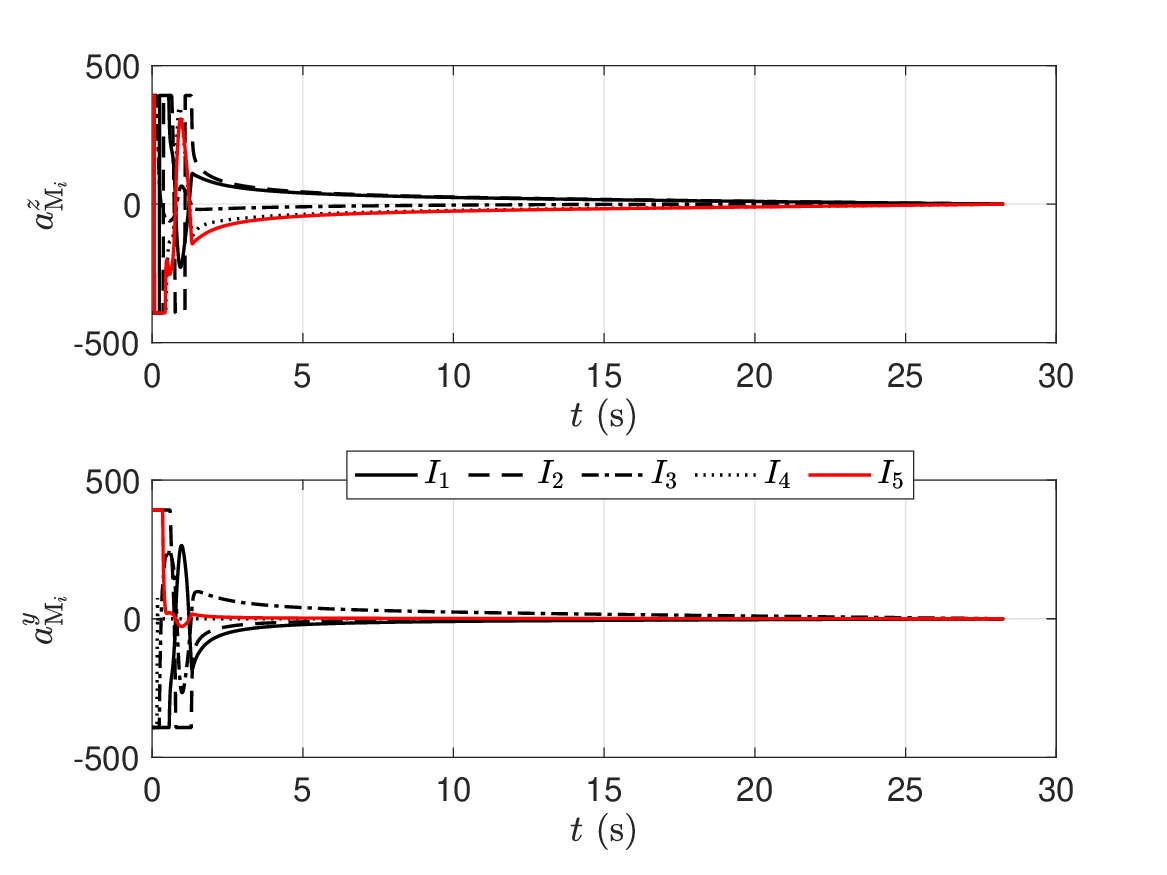}
			\caption{}
			\label{fig:L2am}
		\end{subfigure}
		\hfill
		\begin{subfigure}[t]{0.245\linewidth}
			\centering
			\includegraphics[width=1.1\linewidth]{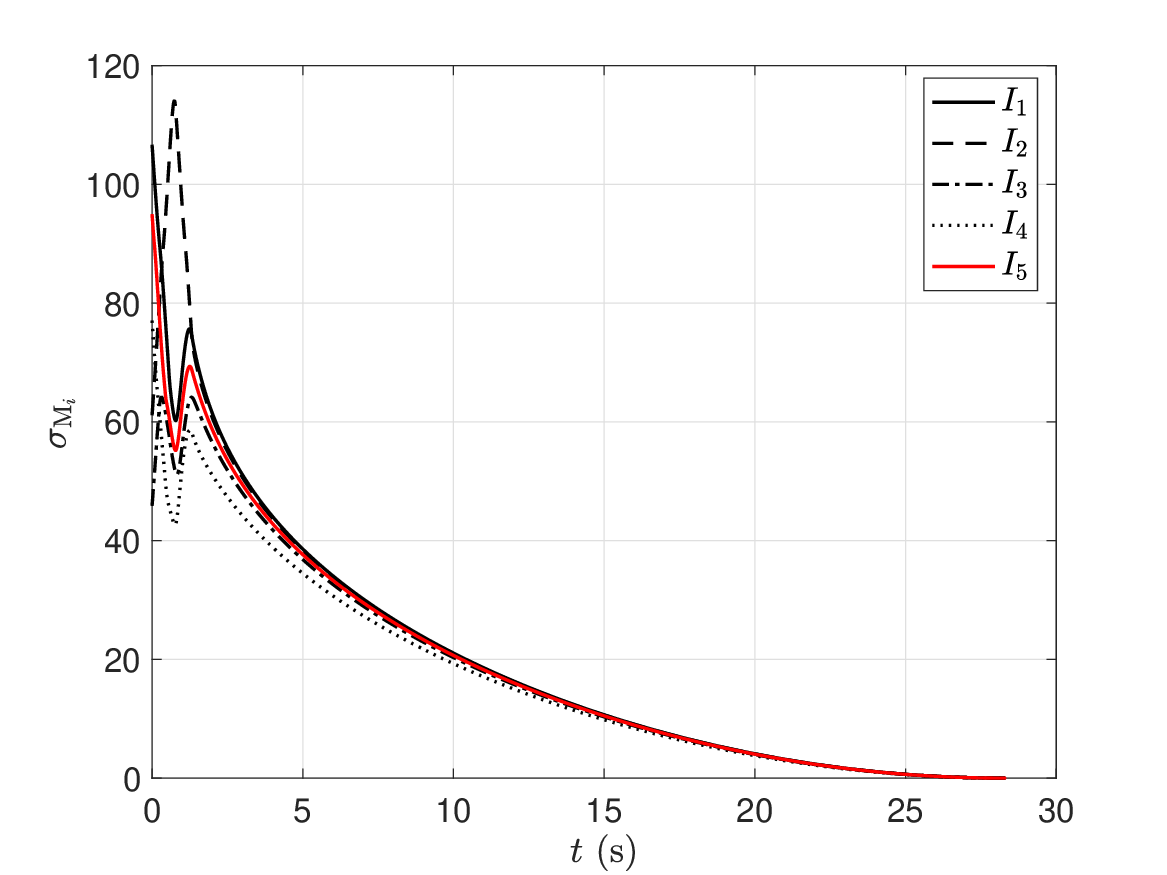}
			\caption{}
			\label{fig:L2sigma}
		\end{subfigure}
		\caption{Cooperative simultaneous target interception (control effort allocation for $\ell=2$ and uncertainty $w_i= 0$).}
		\label{fig:L2}
	\end{figure*}
     \begin{figure*}[h!]
		\begin{subfigure}[t]{0.245\linewidth}
			\centering
			\includegraphics[width=1.1\linewidth]{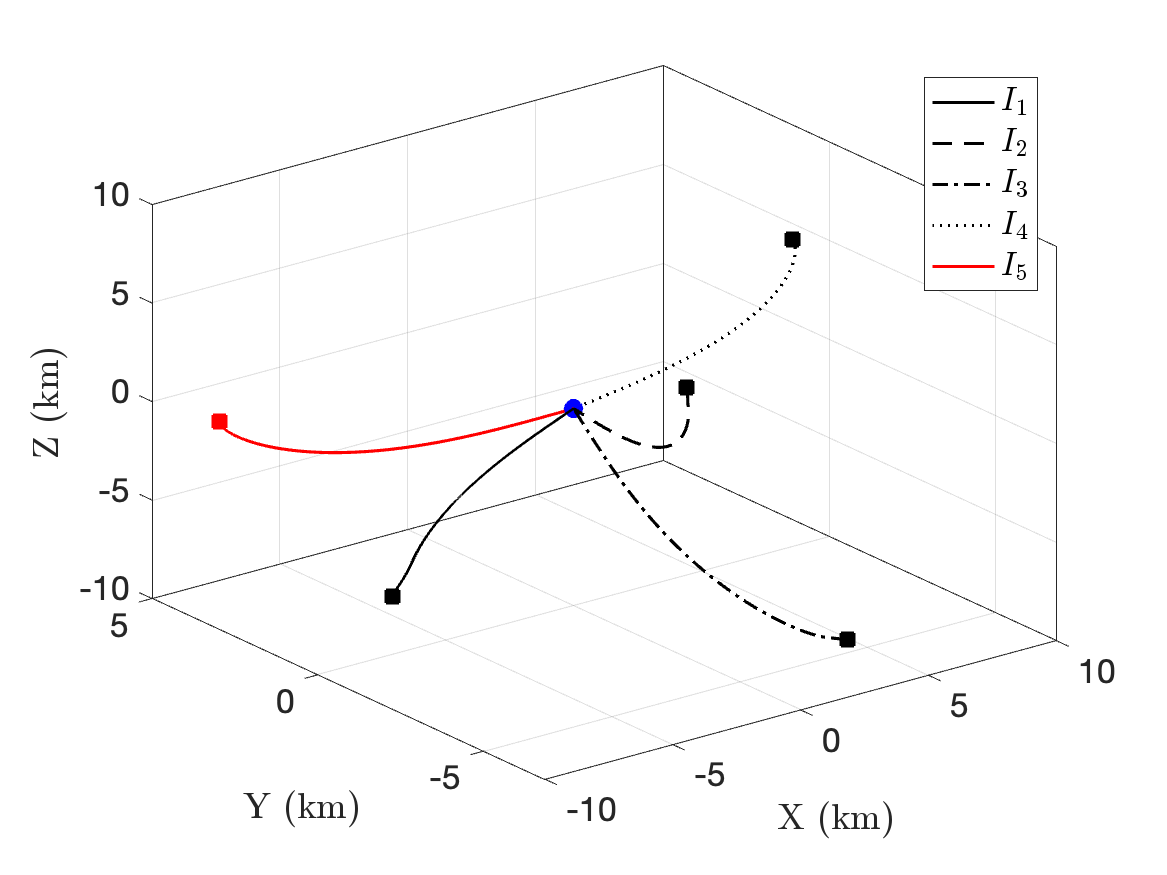}
			\caption{}
			\label{fig:Linftrajectory}
		\end{subfigure}
		\hfill
		\begin{subfigure}[t]{0.245\linewidth}
			\centering
			\includegraphics[width=1.1\linewidth]{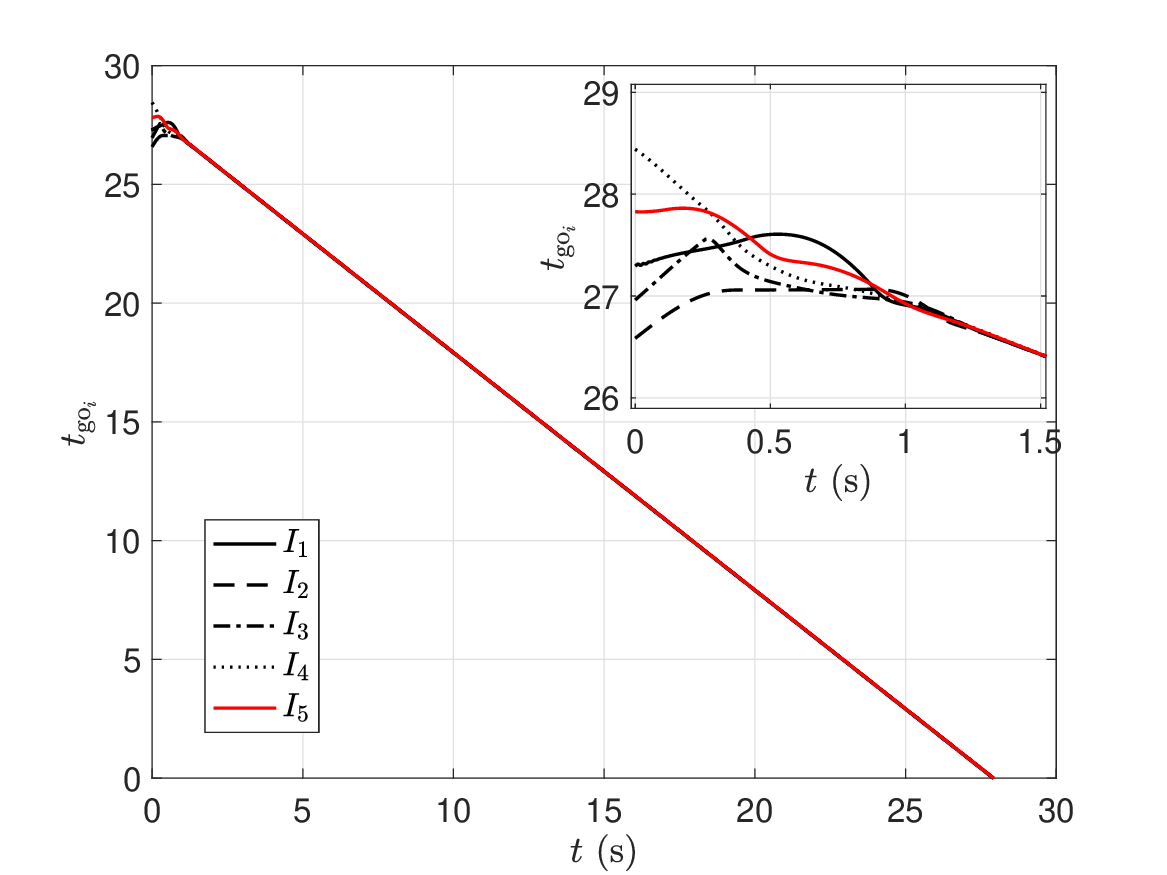}
			\caption{}
			\label{fig:Linftgo}
		\end{subfigure}
		\begin{subfigure}[t]{0.245\linewidth}
			\centering
			\includegraphics[width=1.1\linewidth]{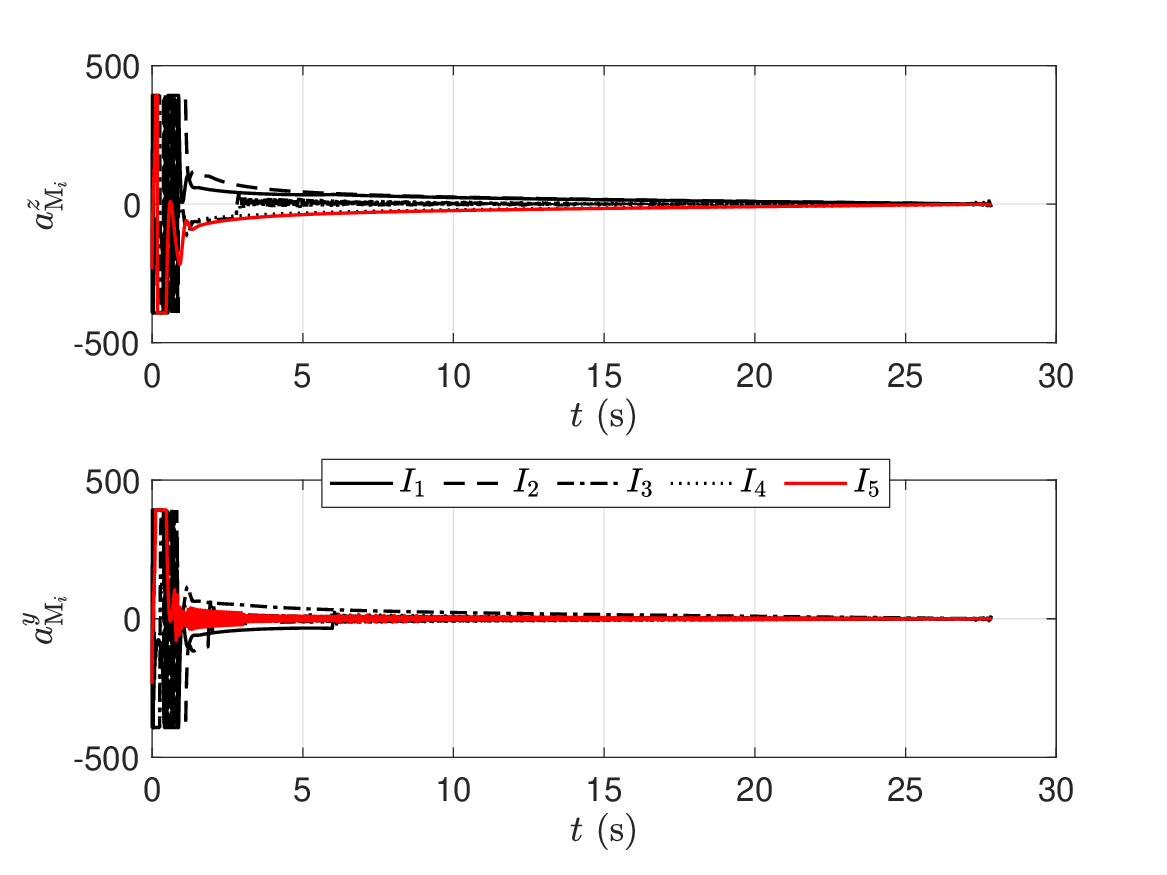}
			\caption{}
			\label{fig:Linfam}
		\end{subfigure}
		\hfill
		\begin{subfigure}[t]{0.245\linewidth}
			\centering
			\includegraphics[width=1.1\linewidth]{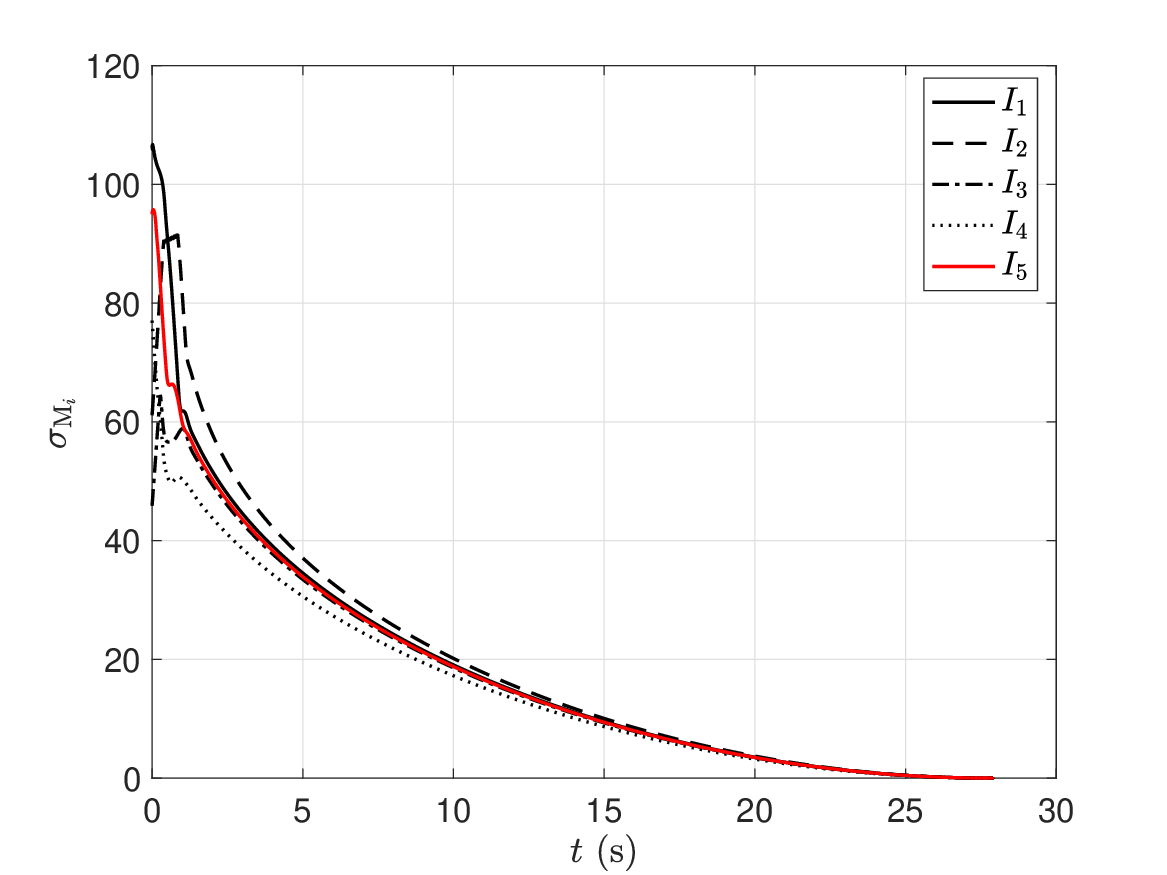}
			\caption{}
			\label{fig:Linfsigma}
		\end{subfigure}
		\caption{Cooperative simultaneous target interception (control effort allocation for $\ell\to\infty$ and uncertainty $w_i= 0$).}
		\label{fig:Linf}
	\end{figure*}
\begin{figure*}[ht!]
	\centering
	\begin{subfigure}[t]{.32\linewidth}
		\centering
		\includegraphics[width=\linewidth]{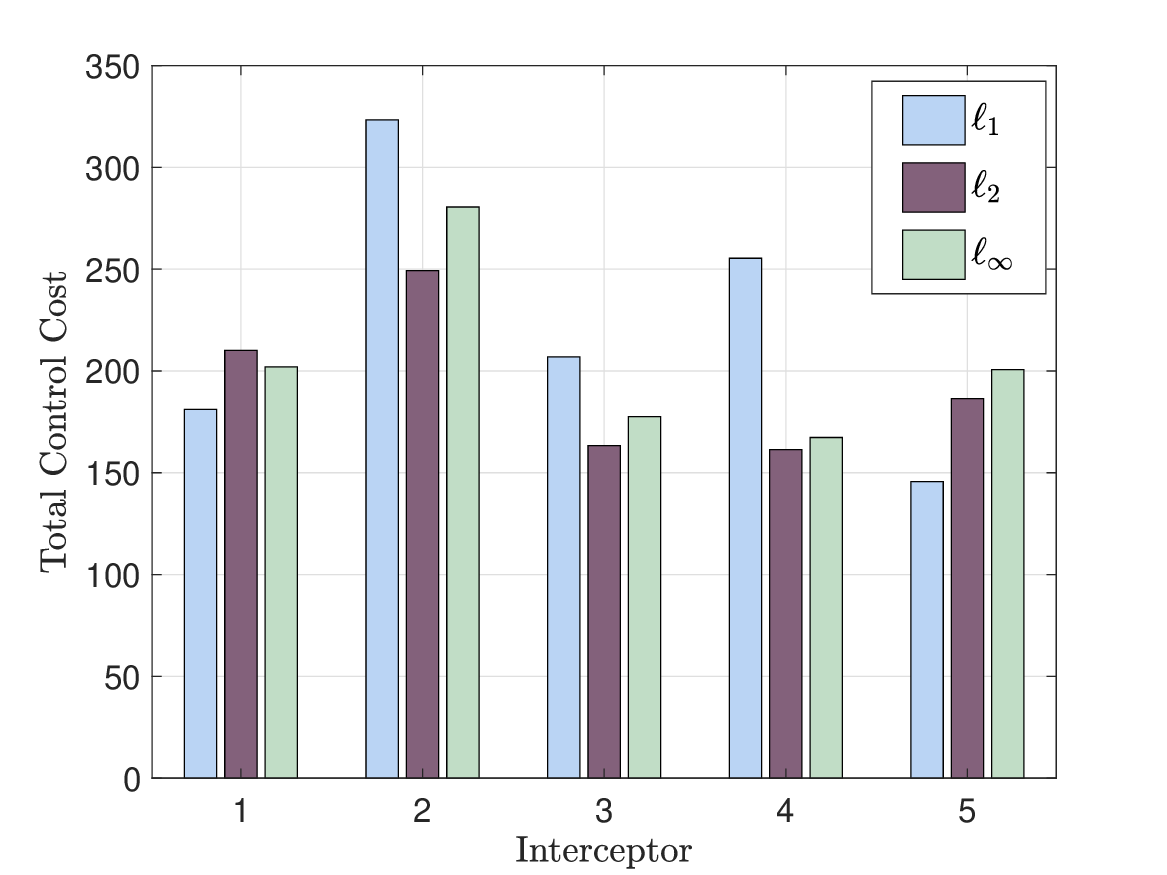}
		\caption{Normalized cost for individual maneuver.}
		\label{fig:trajVariousTimes}
	\end{subfigure}
    \begin{subfigure}[t]{.32\linewidth}
		\begin{tikzpicture}
        {\tiny
			\begin{axis}[
				ybar,
				enlargelimits=0.15,
				legend style={at={(0.5,-0.2)},
					anchor=north,
                    legend columns=-1,
                    font=\footnotesize  
                    },
				ylabel={$\mathcal{J}_i$},
				xtick={1,0.8724,0.9242},
				xticklabels={$\ell_1$,$\ell_2$,$\ell_\infty$},
				width={.95\linewidth}, 
				xtick=data,
				nodes near coords, 
				nodes near coords align={vertical},
                tick label style={font=\footnotesize}, 
		label style={font=\footnotesize},       
				]
				\addplot coordinates {(1,1) (2,0.8724) (3,0.9242)};
			\end{axis}
            }
		\end{tikzpicture}
		\caption{Normalized cost for joint maneuver.}
		\label{fig:JCostCompare}
	\end{subfigure}
    \begin{subfigure}[t]{.32\linewidth}
		\centering
		\includegraphics[width=\linewidth]{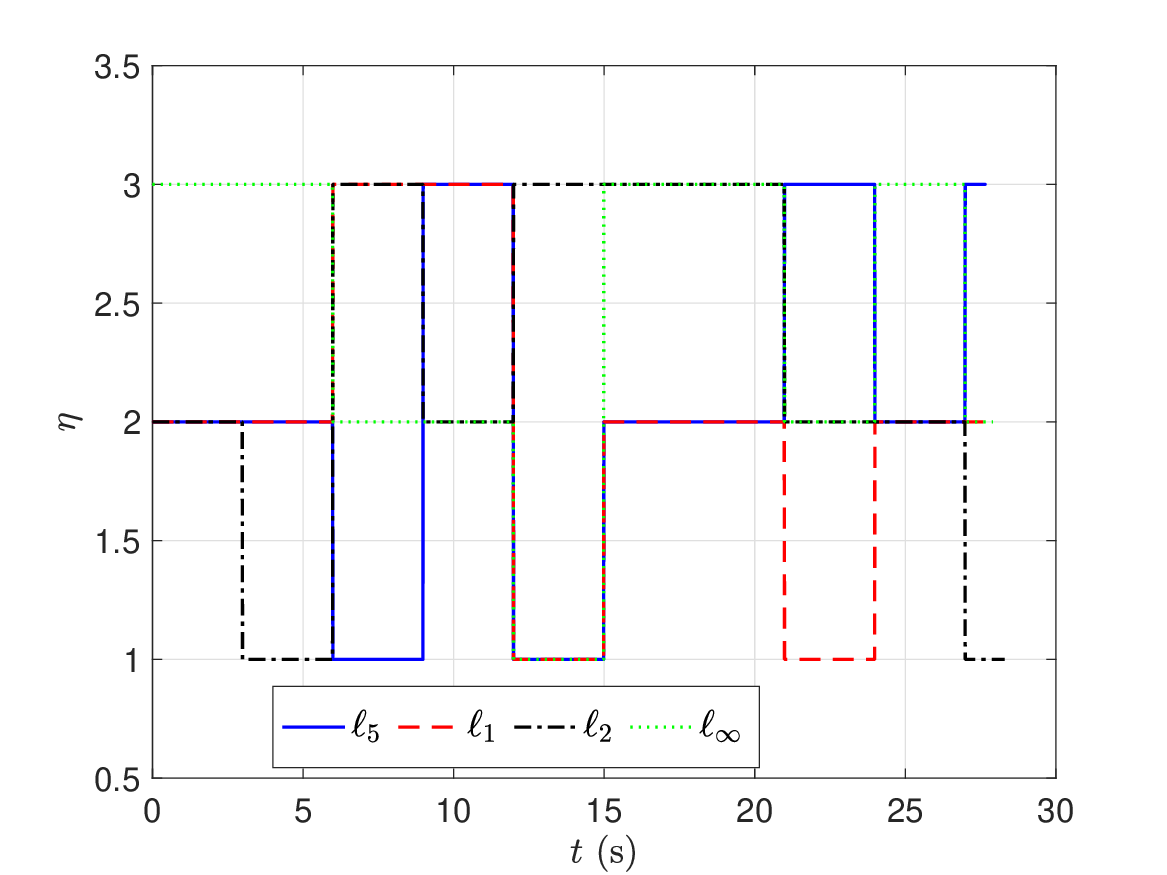}
		\caption{Switching signals.}
    \label{fig:eta}
	\end{subfigure}
	\caption{Comparison of cumulative control effort per agent, and the switching signals.}
	\label{fig:CostCompare}
\end{figure*}

\Cref{fig:L5} depicts a scenario where the time-to-go estimate consists of a lumped uncertainty, sampled from a uniform distribution within the interval $[-3,3]$ s at the beginning of the engagement. The lumped uncertainty is chosen to be an exponential function of $\sigma\M$, that is, $w_i = b_0 e^{-s_0/\sigma\M}$, where $b_0$ is chosen arbitrarily from the set interval and $s_0=8$. It is also assumed that $\|w_i-w_j\|$ is very small because the same estimator used to obtain time-to-go for each interceptor will have fixed characteristics, and any estimator will show convergence to the true value over time. As interceptors attain their desired trajectories (\Cref{fig:L5trajectory}) and agree on their time-to-go (\Cref{fig:L5tgo}), their effective heading angles start to decrease toward zero (\Cref{fig:L5sigma}). During this time, the uncertainties in time-to-go should also decrease and tend toward range over closing speed. Therefore, the effect of uncertainties is only pronounced during the transient phase. This can be seen from \Cref{fig:L5tgo}, where the blue and green line plots in the inset represent the time-to-go profiles when $w_i=0$, and the red and black line plots depict the time-to-go values with uncertainties. Within the predetermined time of consensus $T_s$, the interceptors agree on their time-to-go values regardless of the uncertainties to place the interceptors on the requisite trajectories leading to a cooperative simultaneous interception. In \Cref{fig:L5trajectory}, the red and black markers represent the initial locations of the interceptors, whereas the blue marker is the stationary target. If the uncertainty is persistent, then refinement in the time-to-go estimation method may be necessary and is a separate area of investigation. In the case of \Cref{fig:L5}, the lateral acceleration components are obtained by minimizing $\mathcal{J}_i$ for $\ell=5$. It can be observed from \Cref{fig:L5am} that the lateral acceleration demands subside to zero in the endgame, which is a desirable feature in terminal-phase guidance. One may also notice that the effective heading angles are allowed to be large as seen from \Cref{fig:L5sigma}, which also decreases monotonically to zero as the engagement proceeds.

\Crefrange{fig:L1}{fig:Linf} depict scenarios where $w_i=0$ and control efforts are allocated differently. One may notice that the trajectories are slightly different because the effective heading angle profiles differ. The time-to-go values show different convergence patterns, but they achieve consensus within the chosen $T_s$ regardless of initial engagement configurations. The lateral acceleration components also have relatively larger magnitudes in the transient phase to allow for trajectory correction, and then decay to zero in the endgame. In all scenarios considered, simultaneous interception is consistently achieved at approximately 28 s. This consistency demonstrates the robustness of the proposed method to uncertainties. Furthermore, despite the varying maneuvering demands imposed by different engagement geometries and maneuverability levels, the impact time remains invariant, indicating that the method effectively guarantees a cooperative salvo. For the cases in \Crefrange{fig:L1}{fig:Linf}, a cumulative control effort comparison is shown, where it is observed that the interceptors may utilize different levels of maneuver depending on the method of allocation. In these typical scenarios, $\ell_2$-norm based minimization is found to perform better than other allocations, which is directly related to the power expenditure of the interceptors.
	
	\section{Conclusions}\label{sec:conclusions}
	In this paper, we presented a predefined-time consensus-based cooperative guidance framework for coordinating a swarm of interceptors to achieve simultaneous target interception under switched dynamic networks and uncertainties in time-to-go. The proposed guidance strategy ensures convergence of the interceptors’ time-to-go values to a common value within a user-specified, predefined time, irrespective of the initial engagement, whereas the time of simultaneous interception was decided implicitly during engagement. Such flexibilities in timings eliminate the need for conservative gain selection typically required to ensure prescribed performance, thereby enabling each interceptor to generate feasible trajectories that ensure synchronized arrival at the target. The proposed commands were allocated differently by minimizing different cost functions, and the lateral acceleration components were found to be nonsingular in each case. Ensuring a cooperative salvo against a mobile target under present considerations may be an interesting future work.
	
	\bibliographystyle{IEEEtran}
	\bibliography{references}

\end{document}